\documentclass[
  aps,prd,
  preprint,          %  <<<<<<  toggle: reprint (2-col)  /  preprint (1-col)
  amsmath,amssymb,
  superscriptaddress,
  nofootinbib,
  floatfix,
]{revtex4-2}

\usepackage[T1]{fontenc}
\usepackage{xcolor}
\newcommand{\blue}{\color{blue}}
\usepackage{graphicx}
\usepackage{adjustbox}
\usepackage{hyperref}
\hypersetup{colorlinks=true,linkcolor=blue,citecolor=blue,urlcolor=blue}
\usepackage{amsmath}
\usepackage{slashed}
\usepackage{bm}
\usepackage{array}
\usepackage{booktabs}

\newcommand{\bea}{\begin{eqnarray}}
\newcommand{\eea}{\end{eqnarray}}
\newcommand{\BAN}{\begin{eqnarray*}}
\newcommand{\EAN}{\end{eqnarray*}}

\newcommand{\tr}{{\rm tr}}

\newcommand{\Id}{\mbox{1\hspace{-1.2mm}I}}

\newcommand{\reg}{\text{reg}}
\begin{document}

\title{RG-Invariant Symmetry Ratio for QCD: A Study of
$U(1)_A$ and Chiral Symmetry Restoration}

\author{Ting-Wai Chiu}
\email{twchiu@phys.ntu.edu.tw}
\affiliation{Nuclear Science Division, Lawrence Berkeley National Laboratory, Berkeley, CA 94720, USA}
\affiliation{Department of Physics, National Taiwan University, Taipei 10617, Taiwan}
\affiliation{Institute of Physics, Academia Sinica, Taipei 11529, Taiwan}
\affiliation{Department of Physics, National Taiwan Normal University, Taipei 11677, Taiwan}
\affiliation{Physics Division, National Center for Theoretical Sciences, Taipei 10617, Taiwan}

\author{Tung-Han Hsieh}
\email{thhsieh@gate.sinica.edu.tw}
\affiliation{Research Center for Applied Sciences, Academia Sinica, Taipei 11529, Taiwan}

\begin{abstract}
We introduce a renormalization-group-invariant (RGI), scheme-independent symmetry ratio $\kappa_{AB}$,
for the quantitative characterization of symmetry breaking in QCD. As a first application,
we employ $\kappa_{AB}$ to investigate the relative strength of $SU(2)_L \times SU(2)_R$ chiral symmetry
and $U(1)_A$ axial symmetry breaking in $N_f=2+1+1$ lattice QCD using optimal domain-wall fermions
at the physical point.
Our study covers three lattice spacings and twelve temperatures in the range 164--385~MeV.
We examine three independent symmetry-breaking channels in the nonsinglet sector
with quark-connected correlators:
the $U(1)_A$-sensitive scalar--pseudoscalar channel ($\kappa_{PS}$), probing the $\pi$--$\delta$ system;
the $SU(2)_L \times SU(2)_R$-sensitive vector--axial-vector channel ($\kappa_{VA}$),
probing the $\rho$--$a_1$ system;
and an additional $U(1)_A$-sensitive tensor vector--axial-tensor vector channel ($\kappa_{TX}$),
probing the $\rho_T$--$b_1$ system.
At each lattice spacing we observe a clear hierarchy $\kappa_{PS} > \kappa_{VA} > \kappa_{TX}$.
Since temperature and lattice spacing are not varied independently in the present ensembles,
the lattice-spacing dependence of the hierarchy cannot be isolated, and a controlled continuum
extrapolation is deferred to future work.
\end{abstract}

\maketitle

\section{Introduction}
\label{sec:intro}

The pattern of symmetry realization is a defining feature of any quantum field theory, 
governing its phase structure and the spectrum of its excitations. 
In Quantum Chromodynamics (QCD), the theory of the strong interaction, 
two global symmetries play a pivotal role: the chiral $SU(2)_L \times SU(2)_R$ symmetry of 
the light ($u$, $d$) quark sector, and the $U(1)_A$ axial symmetry.
In the vacuum, the former is spontaneously broken by the chiral condensate $\langle \bar{q} q \rangle$ 
\cite{Nambu:1961tp,Nambu:1961fr}, 
giving mass to nucleons and generating pions as pseudo-Goldstone bosons. 
The latter is explicitly broken by the axial anomaly~\cite{Adler:1969gk,Bell:1969ts,Fujikawa:1979ay}, 
contributing significantly to the mass of the $\eta'$ meson 
\cite{tHooft:1976snw,Witten:1979vv,Veneziano:1979ec}. 
A cornerstone of modern nuclear physics is understanding 
how these broken symmetries behave under extreme conditions of temperature and density, 
such as those realized in heavy-ion collisions or within neutron stars.

The restoration of chiral symmetry is associated with the transition from a hadronic phase 
to a quark-gluon plasma (QGP). For QCD with physical quark masses, this is a smooth crossover occurring 
at a temperature $T_c \sim 156$ MeV 
\cite{Aoki:2006we,Borsanyi:2013bia,HotQCD:2014kol,HotQCD:2018pds,Borsanyi:2020fev}.

A profound and long-standing question is whether 
the effective restoration of the $U(1)_A$ symmetry, linked to the suppression of 
topological gauge fluctuations, coincides with this chiral crossover or occurs at a distinctly 
higher temperature $T_1 > T_c$~\cite{Pisarski:1983ms,Leutwyler:1992yt,Cohen:1996ng,Aoki:2012yj}.
Resolving this hierarchy is essential for a complete understanding 
of the QGP's structure, the nature of the QCD transition, and the validity of effective models. 
Progress on this question via first-principles lattice QCD simulations has been challenging. 
A consistent picture among them has not emerged yet, see e.g., 
refs.~\cite{Cossu:2013uua,Buchoff:2013nra,Brandt:2016daq,Tomiya:2016jwr,Ding:2020xlj,
            Kaczmarek:2023bxb,Aoki:2020noz,Chiu:2023hnm,Gavai:2024mcj}, 
and the recent review \cite{Ding:2026gao}.

The $U(1)_A$ anomaly is particularly sensitive to lattice artifacts, and its clean study requires 
fermion discretizations that preserve chiral symmetry. Lattice studies with chiral fermions 
(domain-wall~\cite{Kaplan:1992bt,Kaplan:1992sg} or overlap~\cite{Neuberger:1997fp,Narayanan:1994gw}) 
have provided crucial insights~\cite{Cossu:2013uua,Buchoff:2013nra,Tomiya:2016jwr,Aoki:2020noz,
Chiu:2023hnm,Gavai:2024mcj}.
Notably, the JLQCD collaboration, using $N_f=2$ M\"obius domain-wall fermions with reweighting 
for overlap fermions at the lattice spacing 0.07 fm, found clear evidence that for $T > 190$~MeV, 
the $U(1)_A$ breaking is consistent with zero within statistical errors~\cite{Aoki:2020noz}.
This is also consistent with studies using $N_f=2+1+1$ optimal domain-wall fermions 
at the physical point and lattice spacing 0.064 fm, where $U(1)_A$ axial symmetry is restored for 
$T \gtrsim 190$~MeV~\cite{Chiu:2023hnm}. However, refs.~\cite{Aoki:2020noz, Chiu:2023hnm}
have not determined the $U(1)_A$ symmetry breaking for $T < 190$~MeV. 
On the other hand, in ref.~\cite{Gavai:2024mcj}, 
using $N_f=2+1$ M\"obius domain-wall fermions at multiple lattice spacings, 
the authors observed that the $U(1)_A$ axial symmetry is not restored for $T \lesssim 186$~MeV. 
To investigate whether any discrepancies between these different studies would occur for $T < 190$~MeV, 
a systematic approach to obtain definite continuum-extrapolated results for physical QCD 
with a chirally symmetric action remains a high-priority goal for the community. 
Addressing this goal requires overcoming two interconnected challenges: first, 
performing controlled continuum extrapolations across the temperature range of interest; 
second, developing a robust, quantitative observable to compare symmetry-breaking strength 
across different channels. 
 
Traditional probes, such as hadron thermal and screening masses 
or the behavior of specific correlation functions at a fixed Euclidean distance, 
can be ambiguous or sensitive to analysis choices. What is needed is a 
\emph{renormalization-group (RG) invariant} measure that integrates spectral information, 
provides a clear normalization, and allows for a direct comparison between 
$SU(2)_L \times SU(2)_R$ and $U(1)_A$ breaking.

In this paper, we address the second of these challenges. The first is deferred to the
dedicated ensembles described in section~\ref{subsec:future}.
First, we introduce a novel, universal diagnostic: 
the \emph{renormalization-group-invariant (RGI) symmetry ratio} $\kappa_{AB}$. 
For two operators $A$ and $B$ related by a symmetry transformation, 
we define $\kappa_{AB} = |\chi_A^\reg - \chi_B^\reg|/(|\chi_A^\reg| + |\chi_B^\reg|)$,
where $\chi_A^\reg$ and $\chi_B^\reg$ denote the corresponding regularized susceptibilities.
For partners of the same sign the definition coincides with
$|(\chi_A^\reg-\chi_B^\reg)/(\chi_A^\reg+\chi_B^\reg)|$. The absolute values keep
$\kappa_{AB}$ within $[0,1]$ for any signs of the partners, with $\kappa_{AB}=1$ if their signs
are opposite. This construct is RG-invariant and scheme-independent for exact symmetry
partners, making it an ideal model-independent probe.

Second, we present the first application of $\kappa_{AB}$ to the problem of symmetry restoration in QCD. 
We perform lattice simulations with $N_f=2+1+1$ optimal domain-wall quarks at the physical point, 
using three lattice spacings and twelve temperatures in the range 164--385~MeV. 
We compute $\kappa_{AB}$ across three distinct symmetry-breaking channels 
in the \emph{nonsinglet} sector: 
the $U(1)_A$-breaking (scalar, pseudoscalar) channel ($\kappa_{PS}$) probing the $\pi$-$\delta$ system; 
the $SU(2)_L\times SU(2)_R$-breaking (vector, axial-vector) channel ($\kappa_{VA}$) 
probing the $\rho$-$a_1$ system; and an additional $U(1)_A$-sensitive channel 
using tensor vector--axial-tensor vector operators ($\kappa_{TX}$) which probes the $\rho_T$-$b_1$ system through 
the $\bar{q}\gamma_4\gamma_k q$ and $\bar{q}\gamma_5\gamma_4\gamma_k q$ currents. 
The inclusion of the tensor vector channel provides a vital cross-check, 
as it probes the $U(1)_A$ anomaly through a different Dirac structure than the scalar channel.

Our key findings are as follows. At finite lattice spacing, we observe a clear ordering: 
$\kappa_{PS} > \kappa_{VA} > \kappa_{TX}$,  
where the breaking in the $U(1)_A$ scalar--pseudoscalar channel is strongest, 
while the $U(1)_A$ tensor vector--axial-tensor vector channel is significantly weaker than the other two channels.
Our ensembles do not share common physical temperatures across lattice spacings, so
temperature and lattice spacing are not varied independently, and a direct
fixed-temperature continuum extrapolation is not possible with the present data. We
therefore present our results as a finite-lattice-spacing study.

The paper is organized as follows. In section~\ref{sec:theory}, we formally define the 
RGI symmetry ratio $\kappa_{AB}$ and detail its theoretical properties. Our lattice setup, 
time-correlation functions, and analysis methodology are described in section~\ref{sec:lattice}. 
Numerical results at finite lattice spacing are presented in section~\ref{sec:results}. 
The implications of our findings and future applications of the $\kappa_{AB}$ framework 
are discussed in section~\ref{sec:discussion}, including the relation of the nonsinglet
$(\pi,\delta)$ splitting to the topological susceptibility (section~\ref{subsec:topology})
and an outline of the work in preparation (section~\ref{subsec:future}).

\section{RGI symmetry ratio}
\label{sec:theory}

We introduce a renormalization-group invariant quantity that quantifies the degree of symmetry 
breaking in a quantum field theory. The construction relies on integrated spectral weights of 
Euclidean correlation functions (susceptibility) for symmetry-related operators. 
Our measure, the \textit{RGI symmetry ratio} $\kappa_{AB}$, 
provides a global, scale-free indicator of symmetry violation and will be applied later 
to $SU(2)_L \times SU(2)_R$ chiral symmetry and $U(1)_A$ symmetry restoration 
in finite-temperature QCD.

A detailed derivation/discussion of the 
UV structure and the renormalization properties of the correlator and the susceptibilty 
is presented in Ref.~\cite{Chiu:2026upk}, which is summarized below.   

\subsection{Correlation functions and symmetry partners}
\label{subsec:correlators}

In a theory with an exact symmetry, correlation functions of operators related by the 
symmetry transformation must be degenerate. 
Consider the Euclidean time correlation function of a local meson operator
$$
O_\Gamma^a(\vec{x},t)=\bar{q}(\vec{x},t)\,\Gamma\,t^a\,q(\vec{x},t),
$$
where $\Gamma$ is a Dirac matrix specifying the quantum numbers of the channel,
$t^a$ (with $a=1,\dots,N_f^2-1$) are the generators of $SU(N_f)$ for flavor nonsinglets,
and $t^0 \equiv \mathbf{1}/\sqrt{2N_f}$ (proportional to the identity) for the flavor singlet,
with the normalization condition $\tr(t^a t^b) = \delta^{ab}/2$ ($a=0,\dots,N_f^2-1$). 
For flavor-singlet operators we adopt the shorthand notation 
$\bar{q} \Gamma q \equiv \bar{q} \Gamma t^0 q$, 
with the factor $t^0$ always implied but suppressed.
For flavor nonsinglet operators, there are $N_f(N_f-1)$ off-diagonal ones with off-diagonal $t^a$
(e.g., $t_1 = \tau_1/2$ and $t_2 = \tau_2/2$ in $SU(2)$)
and $(N_f-1)$ diagonal ones with diagonal $t^a$ 
(e.g., $t_3 = \tau_3/2$ in $SU(2)$).

We use the notation $(x_1,x_2,x_3,x_4)\equiv(x,y,z,t)\equiv(\vec{x},t)$ interchangeably.
With this notation, the $t$-correlator is defined as
\begin{equation}
C_\Gamma(t)=\int d^3 x\;
\bigl[\bigl\langle O_\Gamma^a(\vec{x},t)\,O_\Gamma^{a\dagger}(\mathbf{0},0)\bigr\rangle
- \bigl|\bigl\langle O_\Gamma^a\bigr\rangle\bigr|^2\bigr],
\label{eq:Ct}
\end{equation}
and its spatial counterpart, the $z$-correlator, as
\begin{equation}
C_\Gamma(z)=\int dx\,dy\,dt\;
\bigl[\bigl\langle O_\Gamma^a(x,y,z,t)\,O_\Gamma^{a\dagger}(\mathbf{0},0)\bigr\rangle
- \bigl|\bigl\langle O_\Gamma^a\bigr\rangle\bigr|^2\bigr].
\label{eq:Cz}
\end{equation}
Here $\langle O_\Gamma^a\rangle \equiv v_A$ is the vacuum expectation value (VEV)
of the operator, and the subtraction $-|v_A|^2$ removes the factorized
(constant) piece from the correlator. Here
$O_\Gamma^{a\dagger}=\bar q\,\gamma_4\Gamma^\dagger\gamma_4\,t^a q
=\pm\,O_\Gamma^a$, and the sign makes $C_\Gamma(t)$ and $C_\Gamma(z)$
positive in every channel (section~\ref{sec:lattice}).
Note that for flavor-singlet and diagonal nonsinglet operators,
the full expectation value $\langle O(\vec{x},t)\,O^\dagger(0)\rangle$
receives contributions from both quark-connected and
quark-disconnected Wick contractions.
The terminology ``quark-disconnected'' refers solely to the
quark-line topology. On each gauge configuration $U$, the quark
propagator $D^{-1}(U;x,y)$ is fully dressed by gluon interactions,
and the gauge average $\langle\cdots\rangle_U$ includes the fermion
determinant $\det[D(U)]$ which generates all virtual quark-loop effects.
For most channels $v_A = 0$ by symmetry (except scalar singlet with any quark masses, 
and scalar diagonal nonsinglets with nondegenerate quark masses), 
eqs.~\eqref{eq:Ct}--\eqref{eq:Cz} reduce to $\langle O\,O^\dagger\rangle$.
If a symmetry is exact, the correlators of two partners $A$ and $B$ satisfy $C_A(t)=C_B(t)$ 
for any $t$ and $C_A(z)=C_B(z)$ for any $z$.

Composite operators such as $O_\Gamma^a(x)$ require regularization and renormalization. 
Both $C_\Gamma(t)$ and $C_\Gamma(z)$ contain short-distance singularities. The leading divergence 
behaves as $t^{-3}$ or $z^{-3}$ in the continuum limit as the two operators approach coincidence, 
corresponding to a contact term $\sim a^{-3}$ on the lattice whose precise form is 
regularization-dependent. 
In what follows we concentrate on the $t$-correlator. The extension to the $z$-correlator 
is straightforward.

\subsection{Renormalization and RG invariance}
\label{subsec:renorm}

To quantify the deviation from degeneracy at a specific Euclidean time one may define 
a pointwise ratio
\begin{equation}
\kappa_{AB}(t)=\frac{|C_A(t)-C_B(t)|}{|C_A(t)|+|C_B(t)|},\qquad t\neq0.
\label{eq:kappa_AB_t}
\end{equation}
An analogous ratio $\kappa_{AB}(z)$ for spatial correlators has been employed to study 
$U(1)_A$ and $SU(2)_L\times SU(2)_R$ symmetry patterns in thermal QCD 
with optimal domain-wall fermions~\cite{Chiu:2024jyz,Chiu:2024bqx}.

At fixed $t \neq 0$, the two operators in $C_\Gamma(t)$ are separated
by a nonzero Euclidean distance, so the correlator is free of additive
UV divergences --- no subtraction is required.
The only UV issue is the multiplicative renormalization factor $Z_A^2$
from the anomalous dimension of the composite operator.
This factor cancels in the ratio $\kappa_{AB}(t)$ when $Z_A = Z_B$,
making $\kappa_{AB}(t)$ RG-invariant. 
Additive divergences arise only upon integration over $t$
(the accumulation of the $C(t) \sim 1/t^3$ singularity as $t \to 0$
generates a $\sim 1/a^2$ divergence in the integrated susceptibility),
and their removal is the subject of section~\ref{subsec:integrated}.

\noindent
\textbf{Equality of renormalization constants from symmetry.}
If two operators $O_A$ and $O_B$ are related by a symmetry of the
regularized action, their renormalization constants are equal,
$Z_A = Z_B$~\cite{Chiu:2026upk}. For overlap fermions the
Ginsparg--Wilson relation makes chiral symmetry exact on the lattice at
finite lattice spacing, and $Z_A = Z_B$ holds for all chiral partners, for
any $N_f$ and any quark masses, in any mass-independent scheme.
For domain-wall fermions the same holds in the $N_s\to\infty$ limit.
At finite $N_s$ the residual chiral breaking induces an $O(m_{\rm res})$
splitting between partners, which vanishes exponentially as
$N_s\to\infty$. The equality is unaffected by spontaneous symmetry
breaking or by the $U(1)_A$ anomaly. Moreover, the non-anomalous $SU(2)_A$ cross multiplets tie the
singlet ($s$) and nonsinglet ($ns$) operators of the scalar, pseudoscalar, tensor,
and axial-tensor densities, giving
\begin{equation}
Z_S^{ns} = Z_P^{ns} = Z_S^s = Z_P^s, \qquad
Z_T^{ns} = Z_X^{ns} = Z_T^s = Z_X^s,
\end{equation}
for any $N_f$ and any quark masses, in any mass-independent scheme. 
The derivations are given in Ref.~\cite{Chiu:2026upk}.

For the vector and axial-vector channels the renormalization constants are
finite, and the exact lattice chiral symmetry gives $Z_V = Z_A$. Only this
equality enters $\kappa_{VA}$. The individual value of $Z_V = Z_A$ is
irrelevant to the ratio and is discussed in Ref.~\cite{Chiu:2026upk}.
Note that the singlet vector current
$\bar q t^0 \gamma_\mu q$ and singlet axial-vector current
$\bar q t^0 \gamma_5\gamma_\mu q$ each is invariant under
$SU(2)_L\times SU(2)_R$ (they are flavor singlets) and under $U(1)_A$
(since $\{\gamma_5,\gamma_\mu\}=0$). Thus they are not symmetry partners under either symmetry, 
and are absent from the correlator degeneracy relations
(\ref{eq:C_sigma_pi})-(\ref{eq:C_omega_h1})
for detecting the restoration of $SU(2)_L\times SU(2)_R$ or $U(1)_A$.

These $Z_A=Z_B$ relations ensure that $\kappa_{AB}(t)$ in eq.~\eqref{eq:kappa_AB_t} is
RG-invariant for any pair of operators in the same renormalization
multiplet, as the common factor $Z_A^2 = Z_B^2$ cancels between numerator
and denominator.

\noindent
\textbf{Correlator symmetry relations.}
When a symmetry is effectively restored at temperature $T$, the correlators
of symmetry partners become degenerate: $C_A(t,T) = C_B(t,T)$ for all $t$.
The operator phases are those for which the symmetry implies $C_A=+C_B$, as confirmed by
the degeneracy $\chi_A=\chi_B$ at the reference temperature $T_r$ of
section~\ref{subsec:integrated} (table~\ref{tab:chi_Tr}). The ratio $\kappa_{AB}$ is defined relative to this convention.
The specific partner identifications depend on the number of flavors
through the generators $t^a$ of the flavor group $SU(N_f)$.

In what follows we specialize to $N_f = 2$, where the nonsinglet generators
are $t^a = \tau^a/2$ ($a = 1,2,3$, with $\tau^a$ the Pauli matrices)
and $t^0 = \mathbf{1}/2$ for the singlet.
Effective restoration of $SU(2)_L \times SU(2)_R$ implies
\begin{align}
C^{s}_S(t) &= C^{ns}_P(t)  \Longleftrightarrow  C_\sigma(t) = C_\pi(t),
\label{eq:C_sigma_pi} \\
C^{s}_P(t) &= C^{ns}_S(t)  \Longleftrightarrow  C_\eta(t) = C_\delta(t),
\label{eq:C_eta_delta} \\
C^{s}_{T_k}(t) &= C^{ns}_{X_k}(t)  \Longleftrightarrow  C_{\omega_T}(t) = C_{b_1}(t),
\qquad k=1,2,3, \label{eq:C_omega_b1} \\
C^{s}_{X_k}(t) &= C^{ns}_{T_k}(t)  \Longleftrightarrow  C_{h_{1T}}(t) = C_{\rho_T}(t),
\qquad k=1,2,3, \label{eq:C_h1_rho} \\
C^{ns}_{V_k}(t) &= C^{ns}_{A_k}(t)  \Longleftrightarrow  C_\rho(t) = C_{a_1}(t),
\qquad k=1,2,3, \label{eq:C_rho_a1}
\end{align}
and effective restoration of $U(1)_A$ implies
\begin{align}
C^{ns}_S(t) &= C^{ns}_P(t) \Longleftrightarrow  C_\delta(t) = C_\pi(t),   \label{eq:C_delta_pi}  \\
C^{s}_S(t)  &= C^{s}_P(t)  \Longleftrightarrow  C_\sigma(t) = C_\eta(t),  \label{eq:C_sigma_eta} \\
C^{ns}_{T_k}(t)  &= C^{ns}_{X_k}(t) \Longleftrightarrow  C_{\rho_T}(t) = C_{b_1}(t),
\qquad k=1,2,3, \label{eq:C_rho_b1}  \\
C^{s}_{T_k}(t) &= C^{s}_{X_k}(t) \Longleftrightarrow C_{\omega_T}(t) = C_{h_{1T}}(t),
\qquad k=1,2,3, \label{eq:C_omega_h1}
\end{align}
where the superscripts $s$ and $ns$ denote singlet and nonsinglet channels, respectively.
These partner identifications are the two-flavor chiral multiplets of meson currents
enumerated by Cohen and Ji~\cite{CohenJi:1997}: their eqs.~(6)--(7) give the
$SU(2)_L\times SU(2)_R$ relations~(\ref{eq:C_sigma_pi})--(\ref{eq:C_h1_rho}), their
eq.~(11) the vector--axial-vector relation~(\ref{eq:C_rho_a1}), and their eq.~(8) the
$U(1)_A$ relations~(\ref{eq:C_delta_pi})--(\ref{eq:C_omega_h1}), including the tensor pair
$(\rho_T,b_1)$ used here.
Each pair of (\ref{eq:C_sigma_pi})-(\ref{eq:C_omega_h1})
corresponds to a distinct ratio $\kappa_{AB}(t)$ as defined in eq.~\eqref{eq:kappa_AB_t}.
The notations of meson operators are summarized in appendix~\ref{app:notation}
and table~\ref{tab:meson_notation}.

\subsection{Integrated spectral weights and the RGI symmetry ratio}
\label{subsec:integrated}

Although $\kappa_{AB}(t)$ is a valid probe, its dependence on $t$ and the statistical fluctuations 
at individual times make a channel-wide comparison of symmetry breaking cumbersome. 
A more robust global measure is obtained by integrating over Euclidean time, 
which sums the spectral weight in each channel. 
We therefore define the \textit{bare susceptibility} for channel $\Gamma$ at
temperature $T$ as the integral of the correlator over the thermal circle,
\begin{equation}
\label{eq:chi_bare}
\chi_\Gamma(T) = \int_{0}^{1/T} dt\; C_\Gamma(t,T),
\end{equation}
where $C_\Gamma(t,T)$ is the correlator~\eqref{eq:Ct} at temperature $T$, and
$t=0$ and $t=1/T$ are the same coincident point on the thermal circle. The
integral is regulated by the lattice cutoff $\Lambda = 1/a$. It contains an
additive power divergence $\sim \alpha_\Gamma/(2a^2)$ generated by the
short-distance singularity at the coincident point [eq.~\eqref{eq:chi_UV}
below], which is temperature-independent and is removed by the temperature
subtraction.
Since the VEV subtraction $-\langle O_\Gamma \rangle_T^2$ is already built into the
definition of $C_\Gamma(t,T)$, no additional vacuum subtraction is needed here.
For channels where $v_\Gamma = 0$ (all off-diagonal nonsinglet channels,
and most singlet and diagonal nonsinglet channels), the VEV term is absent and
$C_\Gamma(t,T)$ reduces to $\int d^3x\,\langle O\,O^\dagger\rangle$.
For channels where $v_\Gamma \neq 0$ (the scalar singlet $\sigma$ for any quark masses,
and scalar diagonal nonsinglets for non-degenerate quark masses),
the VEV must be measured on the lattice and subtracted as part of the
correlator construction.

The \textit{regularized susceptibility} is defined as the
difference with respect to a reference temperature $T_r \gg T_c$
at which the symmetry is effectively restored:
\begin{equation}
\label{eq:chi_reg}
\chi_\Gamma^\reg(T;\,T_r) \equiv \chi_\Gamma(T) - \chi_\Gamma(T_r).
\end{equation}
All UV divergences in $\chi_\Gamma(T)$ are independent of temperature
\cite{Dolan:1973qd,Weinberg:1974hy,Bernard:1974bq,Kislinger:1975ab,Landsman:1986uw},
and therefore the additive divergences cancel in $\chi_\Gamma^\reg$, 
rendering it multiplicatively renormalizable. 

\textbf{UV divergence structure.}
The ultraviolet structure of $\chi_A^{\rm bare}$ is derived in detail in 
Ref.~\cite{Chiu:2026upk}. We summarize the results needed here.
In the massless limit the bare susceptibility separates into an additive
power divergence and a multiplicatively renormalized remainder,
\begin{equation}
\chi_\Gamma^{\rm bare}(T, a) = \frac{\alpha_\Gamma}{2 a^2}
+ Z_\Gamma^{2}(\mu, a)\,\chi_\Gamma^R(\mu, T) + O(a),
\label{eq:chi_UV}
\end{equation}
where $\alpha_\Gamma/(2a^2)$ is the leading power divergence from the identity
operator in the operator product expansion (channel-dependent through the
Dirac trace, but temperature-independent), $\chi_\Gamma^R(\mu,T)$ is the
renormalized, ultraviolet-finite susceptibility, and $Z_\Gamma(\mu,a)$ is the
multiplicative renormalization constant of the bilinear $O_\Gamma$. The latter
carries a logarithmic dependence on the lattice spacing, governed by the
anomalous dimension $\gamma_\Gamma$ and resummed by the renormalization group.
For a nonzero quark mass there is one further additive divergence, a
\emph{logarithm} $c_m^\Gamma\,m^2\ln(1/(am))$. This term is chirally even, and
is present even for chirally symmetric fermions and in the continuum.
Exact chiral symmetry forbids all mass-dependent \emph{power} divergences
of the susceptibility. In particular the only dimension-two candidate,
$\propto m/a$, is chirally odd and has an identically vanishing coefficient in 
the Dirac trace~\cite{Chiu:2026upk}.

All of these additive divergences originate from the short-distance region
and are therefore temperature-independent. They cancel exactly in the
temperature subtraction~\eqref{eq:chi_reg}. The multiplicative factor
$Z_\Gamma^{2}$ does not cancel in the subtraction, but cancels in the ratio
$\kappa_{AB}$ when $Z_A = Z_B$, leaving $\kappa_{AB}$ free of all ultraviolet
divergences.

As discussed in Ref.~\cite{Chiu:2026upk}, 
the reference temperature $T_r$ should be chosen sufficiently above $T_c$ such that
the symmetry is effectively restored with $\chi_A(T_r) \simeq \chi_B(T_r)$,  
then the numerator of $\kappa_{AB}$ is independent of $T_r$.
The zeros of $\kappa_{AB}$, signaling degeneracy at $T$, are therefore
common to all admissible choices of $T_r$. 
The denominator serves only for normalization: it carries the same factor 
$Z_A^2 = Z_B^2$ as the numerator, which cancels in the ratio, 
rendering $\kappa_{AB}$ RG-invariant and scheme-independent, and it sets the overall
scale of the ratio, which does depend on $T_r$. Comparisons of
$\kappa_{AB}$ between channels or ensembles are made at a common physical $T_r$.
In this study, we fix the reference temperature at the common physical value
$T_r = 658$~MeV (rather than a fixed $aT_r$), so that the same physical reference is subtracted
at every lattice spacing. At this temperature $\chi_A = \chi_B$ within uncertainties for all
$SU(2)_L \times SU(2)_R$ and $U(1)_A$ nonsinglet channels $PS$, $VA$ and $TX$
(table~\ref{tab:chi_Tr}).

\textbf{Multiplicative renormalization.}
The bilinear renormalizes multiplicatively, 
$O_\Gamma^{\rm bare} = Z_\Gamma(\mu, a)\,O_\Gamma^R$, where $\mu$ is the 
renormalization scale. The logarithmic dependence of $Z_\Gamma$ on the lattice 
spacing $a$ is controlled by the anomalous dimension $\gamma_\Gamma$ of $O_\Gamma$ 
through $\mu\,d\ln Z_\Gamma/d\mu = \gamma_\Gamma$, giving at leading order 
$Z_\Gamma(\mu, a) \sim [\alpha_s(\mu)/\alpha_s(1/a)]^{-\gamma_\Gamma^{(0)}/(2\beta_0)}$.
Thus $Z_\Gamma$ is logarithmically divergent as $a\to 0$ whenever 
$\gamma_\Gamma\neq 0$. For the nonsinglet vector and axial-vector channels 
($\gamma_V^{ns} = \gamma_A^{\rm ns} = 0$) $Z_A^{ns}$ and $Z_V^{ns}$ are finite, and the 
exact lattice chiral symmetry gives $Z_V^{ns} = Z_A^{ns}$. Only this equality is 
needed for $\kappa_{AB}$. 
Correspondingly, the temperature-subtracted susceptibility is related to 
the finite renormalized susceptibility by
\begin{equation}
\chi_\Gamma^\reg(T;\,T_r) = Z_\Gamma^2(\mu,a)\,
\big[\chi_\Gamma^R(\mu,T) - \chi_\Gamma^R(\mu,T_r)\big],
\label{eq:chi_R}
\end{equation}
with the same single-temperature function $\chi_\Gamma^R(\mu,T)$ as in 
eq.~(\ref{eq:chi_UV}), evaluated at the measured and at the reference 
temperature.
For the ratio $\kappa_{AB}$ with $Z_A = Z_B$ (as guaranteed for symmetry partners, 
see section~\ref{subsec:renorm}), 
the common factor $Z_A^2$ cancels and no determination of $Z_A$ is needed.

Because $Z_A=Z_B$ for symmetry partners and $\chi_\Gamma^\reg$ is
multiplicatively renormalizable, we can now construct a global,
RG-invariant measure of symmetry breaking, the
\textit{RGI symmetry ratio}:
\begin{equation}
\boxed{\;
\kappa_{AB}(T) = \frac{|\chi_A^\reg(T;\,T_r) - \chi_B^\reg(T;\,T_r)|}
                      {|\chi_A^\reg(T;\,T_r)| + |\chi_B^\reg(T;\,T_r)|},
\;}
\label{eq:kappa_AB}
\end{equation}
so that $0 \le \kappa_{AB} \le 1$ by the triangle inequality, with $\kappa_{AB}=0$ only if
$\chi_A^\reg=\chi_B^\reg$. For symmetry partners
$\chi_A^\reg$ and $\chi_B^\reg$ carry the same sign throughout our data---positive in the $(P,S)$
channels and negative in $(V,A)$ and $(T,X)$, as follows from the free thermal
coefficients $\tilde c_0^{\Gamma}$ of ref.~\cite{Chiu:2026upk}---and for same-sign
arguments eq.~(\ref{eq:kappa_AB}) coincides identically with
$|(\chi_A^\reg-\chi_B^\reg)/(\chi_A^\reg+\chi_B^\reg)|$. The absolute values keep
$\kappa_{AB}$ within $[0,1]$ for any signs of the partners, without assuming an ordering of
them. If the partners have opposite signs, $\kappa_{AB}=1$, and the splitting is then as
large as the susceptibilities themselves. This case can arise in the $(P,S)$ channel when the valence quarks carry the exact
zero modes, which contribute to $\chi_P$ and $\chi_S$ with opposite signs
(section~\ref{subsec:topology}).
The temperature subtraction removes the additive ultraviolet divergences of
$\chi_\Gamma$. The sign of $\chi_\Gamma^\reg$ that it leaves behind is physical, and
records whether the bare $\chi_\Gamma(T)$ lies above or below $\chi_\Gamma(T_r)$.

\textbf{RG invariance.}
Since $Z_A = Z_B$ for symmetry partners, the common factor $Z_A^2$
cancels between numerator and denominator of $\kappa_{AB}$. The cancellation is
unaffected by the absolute values of eq.~(\ref{eq:kappa_AB}), because $Z_\Gamma^2>0$
and hence $|Z_\Gamma^2\chi_\Gamma^R| = Z_\Gamma^2|\chi_\Gamma^R|$, so that
\begin{equation}
\kappa_{AB}(T) = \frac{|\chi_A^R(T)-\chi_B^R(T)|}{|\chi_A^R(T)|+|\chi_B^R(T)|},
\label{eq:kappa_RGI}
\end{equation}
where $\chi_A^R$ is the finite renormalized susceptibility, 
as defined in eq.~\eqref{eq:chi_R}.
No non-perturbative renormalization (NPR) is needed to evaluate $\kappa_{AB}$.
The parameter $\kappa_{AB}$ provides an intuitive scale:
$\kappa_{AB}=0$ indicates exact degeneracy of the regularized
susceptibilities (the symmetry is manifest), whereas 
$\kappa_{AB}\to 1$ signals a splitting as large as the susceptibilities themselves.

The construction of $\kappa_{AB}$ is general and can be applied to study any
symmetry in quantum field theory for which symmetry partner operators
can be defined.

In the remainder of this paper we employ $\kappa_{AB}$ to investigate the restoration patterns of 
the $U(1)_A$ and $SU(2)_L\times SU(2)_R$ symmetries in high-temperature lattice QCD.

\subsection{Susceptibility symmetry relations}
\label{subsec:chi_sym}

The correlator relations of section~\ref{subsec:renorm},
eqs.~(\ref{eq:C_sigma_pi})--(\ref{eq:C_rho_a1}), integrate over Euclidean time into the
susceptibility equalities collected here for reference. They are used in
section~\ref{subsec:topology} and involve the flavor-singlet channels as well as the
quark-connected ones studied numerically in this work.

For $N_f = 2$ QCD, effective restoration of $SU(2)_L \times SU(2)_R$ chiral symmetry 
(eqs.~(\ref{eq:C_sigma_pi})--(\ref{eq:C_rho_a1})) implies: 
\begin{align}
\chi^{s}_S &= \chi^{ns}_P \Longleftrightarrow \chi_\sigma = \chi_\pi,
\label{eq:chi_sigma_pi} \\
\chi^{s}_P &= \chi^{ns}_S \Longleftrightarrow \chi_\eta = \chi_\delta,
\label{eq:chi_eta_delta} \\
\chi^{s}_{T_k} &= \chi^{ns}_{X_k} \Longleftrightarrow \chi_{\omega_T} = \chi_{b_1},
\qquad k=1,2,3, \label{eq:chi_omega_b1} \\
\chi^{s}_{X_k} &= \chi^{ns}_{T_k} \Longleftrightarrow \chi_{h_{1T}} = \chi_{\rho_T},
\qquad k=1,2,3, \label{eq:chi_h1_rho} \\
\chi^{ns}_{V_k} &= \chi^{ns}_{A_k} \Longleftrightarrow \chi_\rho = \chi_{a_1},
\qquad k=1,2,3, \label{eq:chi_rho_a1}
\end{align}
where the superscripts $s$ and $ns$ denote singlet and nonsinglet channels, respectively.
Recall that the physical $\eta'$ corresponds to the $\eta$ singlet in $N_f=2$ QCD.
The notations of meson operators and their corresponding susceptibilities 
are summarized in appendix~\ref{app:notation} and table~\ref{tab:meson_notation}.

On the other hand, effective restoration of $U(1)_A$ axial symmetry 
(eqs.~(\ref{eq:C_delta_pi})--(\ref{eq:C_omega_h1})) implies:
\begin{align}
\chi^{ns}_S &= \chi^{ns}_P \Longleftrightarrow \chi_\delta = \chi_\pi,  \label{eq:chi_delta_pi}  \\
\chi^{s}_S  &= \chi^{s}_P  \Longleftrightarrow \chi_\sigma = \chi_\eta, \label{eq:chi_sigma_eta} \\
\chi^{ns}_{T_k}  &= \chi^{ns}_{X_k} \Longleftrightarrow  \chi_{\rho_T} = \chi_{b_1},
\qquad k=1,2,3, \label{eq:chi_rho_b1}  \\
\chi^{s}_{T_k} &= \chi^{s}_{X_k} \Longleftrightarrow \chi_{\omega_T} = \chi_{h_{1T}},
\qquad k=1,2,3. \label{eq:chi_omega_h1}
\end{align}
Each equality of (\ref{eq:chi_sigma_pi})-(\ref{eq:chi_omega_h1}) 
corresponds to a distinct RGI symmetry ratio $\kappa_{AB}$
as defined in eq.~\eqref{eq:kappa_AB}.
Since $Z_A = Z_B$ for all these symmetry pairs
(established in section~\ref{subsec:renorm}),
\emph{any} ratio $\kappa_{AB}$ involving operators from the same renormalization
constant multiplet is RG-invariant. This includes the nonsinglet ratios
$\kappa_{PS}$, $\kappa_{VA}$, $\kappa_{TX}$ studied in this work, as well as
singlet-involved ratios such as $\kappa_{\sigma,\pi}$, $\kappa_{\eta,\delta}$,
$\kappa_{\omega_T,b_1}$, $\kappa_{h_{1T},\rho_T}$, $\kappa_{\sigma, \eta}$ and $\kappa_{\omega_T,h_{1T}}$.
Note that the equalities (\ref{eq:chi_sigma_pi})-(\ref{eq:chi_omega_h1}) hold for the
bare, the regularized (temperature subtracted), and the renormalized susceptibilities 
respectively, as well as for prescriptions including/excluding the time slice $t=0$ 
as defined in (\ref{eq:chi_bare_lat_all}) and (\ref{eq:chi_bare_lat}). 

Despite this formal RG invariance, practical extraction of
$\kappa_{AB}$ for channels involving flavor-singlet operators is challenging
in current lattice simulations.
Singlet correlators receive contributions from quark-disconnected
Wick contractions, which are computationally demanding and statistically
noisy. For the scalar singlet $\sigma \sim \bar{q}q$, the nonvanishing
VEV $v_\sigma = \langle\bar{q}q\rangle_T$ must additionally be measured
and subtracted at both $T$ and $T_r$.
These computational difficulties are the main obstacles. The
renormalization of singlet channels is no more complicated than
that of nonsinglet channels, since $Z_A = Z_B$ for all symmetry
pairs (section~\ref{subsec:renorm}).

Given these considerations, the present study focuses on the
quark-connected, nonsinglet channels where the RG‑invariant ratio $\kappa_{AB}$ rests on the firmest
theoretical and numerical ground:
\begin{itemize}
    \item $\kappa_{PS}$ ($\delta$-$\pi$) for $U(1)_A$,
    \item $\kappa_{VA}$ ($\rho$-$a_1$) for $SU(2)_L \times SU(2)_R$,
    \item $\kappa_{TX}$ ($\rho_T$-$b_1$) for $U(1)_A$ through a different Dirac structure.
\end{itemize}
These choices provide a clean, theoretically unambiguous, and statistically precise set of observables.
The three channels are compared in section~\ref{sec:results}.

\noindent
\textbf{Remark on terminology.} Throughout this work, 
``restoration'' of a symmetry always refers to \emph{effective restoration} --- i.e., 
the corresponding RG-invariant ratio $\kappa_{AB}$ becomes consistent with zero 
within statistical uncertainties. 
Exact restoration, which would require vanishing quark masses for $SU(2)_L\times SU(2)_R$ 
and additionally the complete suppression of the $U(1)_A$ anomaly for $U(1)_A$ restoration, 
is neither achievable in physical QCD nor required for the validity of our conclusions. 
This operational definition is implicit in all subsequent uses of 
``degeneracy'', ``restoration'', and ``$\kappa_{AB}=0$''.
Note that $\kappa_{AB}=0$ means the regularized susceptibilities
$\chi_A^\reg$ and $\chi_B^\reg$ are equal at temperature $T$,
i.e., channel $A$ and channel $B$ carry the same integrated
spectral weight above the restored baseline $T_r$.

\noindent
\textbf{Notation.}
Throughout this paper the two susceptibilities are always distinguished
explicitly: $\chi_\Gamma(T)$ denotes the bare susceptibility~\eqref{eq:chi_bare},
and $\chi_\Gamma^\reg$ the regularized one~\eqref{eq:chi_reg}, whose arguments
$(T;\,T_r)$ are suppressed only where the reference temperature is unambiguous.
The RGI symmetry ratios $\kappa_{AB}$ are built from $\chi_\Gamma^\reg$, whereas
the identities of section~\ref{subsec:topology} involve the bare $\chi_\Gamma$.

\section{Lattice setup}
\label{sec:lattice}

We generate gauge ensembles using hybrid Monte Carlo (HMC) simulations of lattice QCD with $N_f = 2+1+1$ 
optimal domain-wall quarks~\cite{Chiu:2002ir,Chiu:2015sea} at the physical point. 
The simulations are performed on $32^3 \times (16,12,10,8,6,4)$ lattices with the 
plaquette gauge action~\cite{Wilson:1974sk} at three values of $\beta = 6/g^2 = (6.15, 6.18, 6.20)$, 
corresponding to lattice spacings $a \simeq (0.075,0.069,0.064)$~fm. 
These ensembles are produced with the same 
actions~\cite{Chiu:2011bm,Chen:2014hyy} and algorithms as their counterparts on larger 
$64^3 \times (20,16,12,10,8,6)$ lattices~\cite{Chen:2022fid}, but at one-eighth of the spatial volume. 
Simulations are carried out on GPU clusters equipped with various NVIDIA GPUs.

After initial thermalization, gauge configurations are sampled and distributed among 16–32 independent 
simulation units, each performing a separate HMC stream. In each stream, one configuration is sampled 
every five trajectories. All sampled configurations from all streams are combined to obtain the final ensemble. 
Lattice parameters and statistics for the meson $t$-correlator calculations are listed in 
table~\ref{tab:ensembles}. The temperatures covered range from $\sim 164$ to $385$~MeV, 
all above the pseudocritical temperature $T_c \sim 156$~MeV.

\begin{table*}[tbp]
\centering
\setlength{\tabcolsep}{3pt}
\vspace{-2mm}
\begin{adjustbox}{max width=\linewidth}
\begin{tabular}{|ccccccccc|}
\hline 
$\beta$ & $a$[fm] & $N_x$ & $N_t$ & $T$[MeV] & $N_{\rm confs}$ & $(m_{u/d}a)_{\rm res}$ & $(m_s a)_{\rm res}$ & $(m_c a)_{\rm res}$ \\
\hline
6.15 & 0.075 & 32 & 16 & 164 & 359 & $6.2(5){\times}10^{-5}$ & $3.1(4){\times}10^{-5}$ & $1.1(7){\times}10^{-5}$\\
6.18 & 0.069 & 32 & 16 & 179 & 337 & $4.9(7){\times}10^{-5}$ & $5.8(1.7){\times}10^{-5}$ & $2.7(9){\times}10^{-5}$\\
6.20 & 0.064 & 32 & 16 & 192 & 588 & $3.5(4){\times}10^{-5}$ & $2.3(2){\times}10^{-5}$ & $6.5(1.1){\times}10^{-6}$\\
6.15 & 0.075 & 32 & 12 & 219 & 409 & $9.0(1.4){\times}10^{-5}$ & $6.0(8){\times}10^{-5}$ & $1.5(3){\times}10^{-6}$\\
6.18 & 0.069 & 32 & 12 & 238 & 781 & $1.9(2){\times}10^{-5}$ & $1.6(1){\times}10^{-5}$ & $3.8(5){\times}10^{-6}$\\
6.20 & 0.064 & 32 & 12 & 257 & 514 & $1.8(6){\times}10^{-5}$ & $1.6(6){\times}10^{-5}$ & $9.4(4.2){\times}10^{-6}$\\
6.15 & 0.075 & 32 & 10 & 263 & 496 & $2.4(4){\times}10^{-5}$ & $2.0(3){\times}10^{-5}$ & $7.5(2.6){\times}10^{-6}$\\
6.18 & 0.069 & 32 & 10 & 286 & 377 & $2.4(8){\times}10^{-5}$ & $2.2(7){\times}10^{-5}$ & $9.7(3.9){\times}10^{-6}$\\
6.20 & 0.064 & 32 & 10 & 308 & 481 & $5.8(1.5){\times}10^{-6}$ & $5.1(1.2){\times}10^{-6}$ & $1.4(2){\times}10^{-6}$\\
6.15 & 0.075 & 32 & 8 & 329 & 640 & $3.5(8){\times}10^{-5}$ & $2.9(6){\times}10^{-5}$ & $1.2(3){\times}10^{-5}$\\
6.18 & 0.069 & 32 & 8 & 357 & 302 & $1.3(7){\times}10^{-5}$ & $1.2(6){\times}10^{-5}$ & $4.9(2.2){\times}10^{-6}$\\
6.20 & 0.064 & 32 & 8 & 385 & 468 & $6.3(1.9){\times}10^{-6}$ & $6.0(1.8){\times}10^{-6}$ & $3.0(9){\times}10^{-6}$\\
6.15 & 0.075 & 32 & 6 & 438 & 602 & $1.1(3){\times}10^{-5}$ & $1.0(3){\times}10^{-5}$ & $4.0(8){\times}10^{-6}$\\
6.18 & 0.069 & 32 & 6 & 477 & 560 & $6.3(1.5){\times}10^{-6}$ & $6.2(1.5){\times}10^{-6}$ & $5.1(1.5){\times}10^{-6}$\\
6.20 & 0.064 & 32 & 6 & 513 & 431 & $5.8(2.4){\times}10^{-6}$ & $5.6(2.4){\times}10^{-6}$ & $4.0(1.4){\times}10^{-6}$\\
6.15 & 0.075 & 32 & 4 & 658 & 413 & $4.5(1){\times}10^{-6}$ & $4.5(1){\times}10^{-6}$ & $9.4(3){\times}10^{-7}$\\
6.18 & 0.069 & 32 & 4 & 715 & 763 & $1.0(1){\times}10^{-6}$ & $1.0(1){\times}10^{-6}$ & $1.0(1){\times}10^{-6}$\\
6.20 & 0.064 & 32 & 4 & 770 & 991 & $1.2(2){\times}10^{-6}$ & $1.2(2){\times}10^{-6}$ & $1.2(2){\times}10^{-6}$\\
\hline
\end{tabular}
\end{adjustbox}
\caption{Lattice parameters and statistics of the eighteen gauge ensembles used in this work.
The twelve ensembles with $N_t=16,12,10,8$ provide the temperatures $164$--$385$~MeV at which
$\kappa_{AB}$ is measured. Those with $N_t=6$ and $N_t=4$ enter only through the reference
temperature, either directly ($T_r=\hbar c/(4a)$) or through the interpolation of the
reference susceptibilities described in section~\ref{sec:lattice}.
The last three columns give the residual masses of $u/d$, $s$, and $c$ quarks~\cite{Chen:2012jya}.
Temperatures are obtained from $T=\hbar c/(N_t a)$ with the nominal spacings
$a=(0.0750,0.0690,0.0641)$~fm, each of which lies within the uncertainty of the
determination quoted in table~\ref{tab:a_qmass}. The same values are used throughout the
analysis, so that in particular $T_r=\hbar c/(4a)=657.8$~MeV at $\beta=6.15$, quoted as
$658$~MeV below.}
\label{tab:ensembles}
\end{table*}

Lattice spacings and quark masses ($u/d$, $s$, $c$) are determined on $32^3 \times 64$ lattices 
with $\{460, 636, 726\}$ configurations for $\beta = \{6.15,6.18,6.20\}$, respectively. 
The lattice spacing is fixed using the Wilson flow~\cite{Narayanan:2006rf,Luscher:2010iy} 
with the condition $\{t^2\langle E(t)\rangle\}|_{t=t_0}=0.3$ and input 
$\sqrt{t_0}=0.1416(8)$~fm~\cite{Bazavov:2015yea}. The resulting spacings are listed in table~\ref{tab:a_qmass}. 
Physical quark masses are obtained by tuning the bare masses so that the lowest-lying states extracted 
from time-correlation functions of the meson operators 
$\{\bar u\gamma_5 d, \bar s\gamma_i s, \bar c\gamma_i c\}$ 
agree with the physical masses of $\pi^\pm(140)$, $\phi(1020)$, and $J/\psi(3097)$. 
The tuned bare quark masses are also given in table~\ref{tab:a_qmass}.

\begin{table*}[tbp]
\centering
\setlength{\tabcolsep}{4pt}
\vspace{2mm}
\begin{adjustbox}{max width=\linewidth}
\begin{tabular}{|ccccc|}
\hline
$\beta$ & $a$[fm] & $m_{u/d}a$ & $m_s a$ & $m_c a$ \\
\hline
6.15 & 0.0751(5) & 0.00200 & 0.064 & 0.705 \\
6.18 & 0.0690(5) & 0.00180 & 0.058 & 0.626 \\
6.20 & 0.0641(4) & 0.00125 & 0.040 & 0.550 \\
\hline
\end{tabular}
\end{adjustbox}
\caption{Lattice spacings and bare quark masses for $N_f=2+1+1$ lattice QCD with optimal domain-wall quarks 
at the physical point.}
\label{tab:a_qmass}
\end{table*}

Chiral symmetry breaking due to the finite extent $N_s = 16$ in the fifth dimension is quantified 
by the residual masses of each quark flavor~\cite{Chen:2012jya}, listed in the last three columns 
of table~\ref{tab:ensembles}. These residual masses are less than $(4.5\%,\;0.1\%,\;0.005\%)$ 
of the corresponding bare masses for $(u/d,s,c)$ quarks, translating to less than 
$(0.2,\;0.1,\;0.06)$~MeV/$c^2$, respectively. This confirms that chiral symmetry is well preserved 
and that the effective 4D Dirac operator for optimal domain-wall fermions remains accurate for 
both light and heavy quarks. Consequently, hadronic observables (such as meson correlators) 
can be computed with high precision, with uncertainties dominated by statistics and other systematics.

We now summarize the notations and conventions used in this work.

The correlation function of off-diagonal flavor-nonsinglet meson interpolator 
$\bar q_1 \Gamma q_2$ (e.g., $\bar u \Gamma d$) on a lattice with $(N_x, N_y, N_z, N_t)$ sites is computed as
\begin{equation}
\label{eq:C_Gamma}
C_\Gamma(x) = \bigl\langle (\bar q_1 \Gamma q_2)_x (\bar q_1 \Gamma q_2)_0^\dagger \bigr\rangle
           = -\Bigl\langle \operatorname{tr} \bigl[ \bar\Gamma\, (D_c + m_1)^{-1}_{0,x} \,
                                            \Gamma\, (D_c + m_2)^{-1}_{x,0} \bigr] \Bigr\rangle_{\text{confs}},
\end{equation}
where $\bar\Gamma \equiv \gamma_4 \Gamma^\dagger \gamma_4 = \pm\Gamma$, $(D_c + m_q)^{-1}$ denotes the valence quark propagator with mass $m_q$ in lattice QCD 
with exact chiral symmetry~\cite{Chiu:1998eu}, $\operatorname{tr}$ is the trace over color and Dirac indices, 
and $\langle\cdots\rangle_{\text{confs}}$ denotes the average over gauge configurations. 
Here $x = (x_1, x_2, x_3, x_4) = (x, y, z, t)$. The sign of $\bar\Gamma$ relative to $\Gamma$, 
together with the minus sign of the fermion loop, makes $C_\Gamma(t)>0$ in every channel, 
consistent with the convention of section~\ref{subsec:renorm}. 
The temporal correlator is defined as
\begin{equation}
\label{eq:C_Gamma_t}
C_\Gamma(t,T) = \sum_{x_1, x_2, x_3} C_{\Gamma}(x),
\end{equation}
where $T = 1/(N_t a)$ is the temperature.
On the lattice the bare susceptibility~(\ref{eq:chi_bare}) is the sum over all
time slices,
\begin{equation}
\label{eq:chi_bare_lat_all}
\chi_\Gamma(T) = \sum_{t=0}^{N_t-1} C_\Gamma(t,T).
\end{equation}
An equally admissible prescription 
omits the $t=0$ time slice, the contact term of the two coincident operators,  
\begin{equation}
\label{eq:chi_bare_lat}
\chi_\Gamma(T) = \sum_{t=1}^{N_t-1} C_\Gamma(t,T).
\end{equation}
Both are legitimate. The additive divergences of either originate in the
zero-temperature propagator at short distances and are therefore
temperature-independent, so they cancel in the temperature
subtraction~(\ref{eq:chi_reg}). The two definitions carry the same
multiplicative renormalization and yield the same $\kappa_{AB}$ in the
continuum limit. The full renormalization analysis, establishing the
equivalence of the two prescriptions channel by channel, is given in
Ref.~\cite{Chiu:2026upk}. In the present work all susceptibilities are
computed with the contact-excluded definition~(\ref{eq:chi_bare_lat}). The
corresponding results for the contact-included
definition~(\ref{eq:chi_bare_lat_all}) will be presented elsewhere, as a
numerical check that both prescriptions give a consistent picture in the
continuum limit.

The symmetry relations among susceptibilities established in
Sec.~\ref{subsec:renorm}, eqs.~(\ref{eq:chi_sigma_pi})--(\ref{eq:chi_omega_h1}),
hold for both prescriptions. They follow by integrating the corresponding
correlator relations~(\ref{eq:C_sigma_pi})--(\ref{eq:C_omega_h1}), which hold
locally at each time slice $t$, including $t=0$. Summing over any set of time slices,
with or without the contact term, therefore preserves the equalities. 

In this study we focus on the off-diagonal flavor-nonsinglet operator $\bar u \Gamma d/\sqrt{2}$, with
$$
\Gamma = \{\Id, \gamma_5, \gamma_k, \gamma_5\gamma_k, \gamma_4\gamma_k, \gamma_5\gamma_4\gamma_k, k=1,2,3\},
$$
corresponding to the scalar (S), pseudoscalar (P), vector (V), axial-vector (A), tensor-vector (T), 
and axial-tensor-vector (X) channels, respectively.

Thanks to $S_3$ symmetry, the correlators satisfy 
$C_{V_1}=C_{V_2}=C_{V_3}$, $C_{A_1}=C_{A_2}=C_{A_3}$, $C_{T_1}=C_{T_2}=C_{T_3}$, and $C_{X_1}=C_{X_2}=C_{X_3}$. 
To improve statistics, we average over the three spatial components for each channel, e.g.,
$$
C_V(t,T) = \frac{1}{3}\sum_{k=1}^3 C_{V_k}(t,T),
$$
and similarly for $A$, $T$ and $X$.

These averaged correlators $C_\Gamma(t,T)$ are used to compute the regularized susceptibility 
of off-diagonal flavor nonsinglet mesons,
\begin{equation}
\label{eq:chi_reg_lat}
\chi_\Gamma^\reg(T;\,T_r) = \sum_{t=1}^{N_t-1} [ C_\Gamma(t,T) - C_\Gamma(t,T_r)],  
\end{equation}
where $T_r = 658$~MeV is a common physical reference temperature, well above $T_c$, at which the
chiral symmetries are highly restored. At $\beta=6.15$ it coincides with $aT_r=1/4$, and on the finer
lattices $C_\Gamma(t,T_r)$ is obtained by log-linear interpolation in $N_t$ between the
$N_t=6$ and $N_t=4$ ensembles of table~\ref{tab:ensembles}.
The RGI symmetry ratios $\kappa_{AB}$ defined in eq.~(\ref{eq:kappa_AB}) 
are then computed for the channel pairs $VA$, $PS$, and $TX$.
Here $\kappa_{VA}$ probes $SU(2)_L\times SU(2)_R$ chiral symmetry restoration, 
while $\kappa_{PS}$ and $\kappa_{TX}$ probe $U(1)_A$ axial symmetry restoration.

Statistical uncertainties on $\chi_\Gamma(T)$, $\chi_\Gamma^\reg$ and $\kappa_{AB}(T)$ are estimated using the 
jackknife method with binning. The central values are computed from the full ensemble, 
and the jackknife variance is evaluated for several bin sizes (typically 5 to 15 configurations). 
The quoted error corresponds to the bin size at which the estimate saturates, 
which accounts for residual autocorrelations between successive configurations.
A crucial feature of this analysis is that $\chi_A^\reg$ and $\chi_B^\reg$ for symmetry partners
are computed from the \emph{same} gauge configurations -- differing only in the Dirac 
structure $\Gamma$ applied to the same quark propagator -- so their fluctuations are 
highly correlated (correlation coefficient $\rho \sim 0.9999$). On each jackknife sample, 
both $\chi_A^\reg$ and $\chi_B^\reg$ are evaluated together and the ratio
$\kappa_{AB}$ of eq.~(\ref{eq:kappa_AB}) is formed within the sample.
The jackknife variance therefore captures this correlation automatically, yielding an 
error on $\kappa_{AB}$ that is substantially smaller than would be obtained from naive 
(uncorrelated) propagation of the individual errors on $\chi_A^\reg$ and $\chi_B^\reg$. 
This is the standard mechanism by which ratios and differences of correlated lattice 
observables (mass splittings between nearly degenerate states, effective mass ratios, 
step-scaling functions) achieve much higher precision than individual measurements.

\section{Results}
\label{sec:results}

\begin{figure*}[tbp]
\centering
  \includegraphics[width=\linewidth,clip=true]{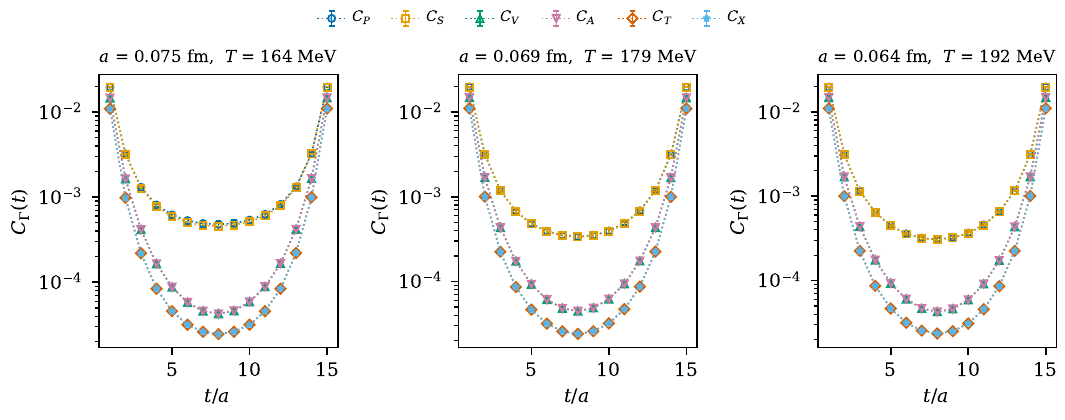}
\caption{\label{fig:Ct_ud} $t$-correlators $C_\Gamma(t)$ of the nonsinglet operators
         $\bar u \Gamma t^a d$ at the three lowest temperatures, including the factor $1/2$ from
         the flavor normalization $\tr(t^at^b)=\delta^{ab}/2$ (tables~\ref{tab:Ct_ud_b615}--\ref{tab:Ct_ud_b620}).
         The dotted lines connecting the data points in each channel are shown only to guide the eye.}
\end{figure*}

We begin by analyzing the temporal correlators $C_\Gamma(t)$ of the $\bar{u}\Gamma d$ bilinears at 
the three lowest temperatures, $T = (164,179,192)$~MeV, displayed in figure~\ref{fig:Ct_ud}. 
Our focus is on the degeneracy patterns among the $(V,A)$, $(P,S)$, and $(T,X)$ channels, 
which reflect the restoration of $SU(2)_L \times SU(2)_R$ chiral symmetry (probed by $V$-$A$ degeneracy) 
and $U(1)_A$ axial symmetry (probed by $P$-$S$ and $T$-$X$ degeneracy).

As shown in figure~\ref{fig:Ct_ud}, the correlators follow the ordering
$$
C_P(t) \gtrsim C_S(t) > C_V(t) \gtrsim C_A(t) > C_T(t) \gtrsim C_X(t)
$$
at all temperatures. This hierarchy corresponds to the ordering of meson thermal masses,
$$
m_P \lesssim m_S < m_V \lesssim m_A < m_T \lesssim m_X ,
$$
which remains consistent across the full temperature range studied (164–385~MeV).

\begin{figure*}[tbp]
\centering
  \includegraphics[width=\linewidth,clip=true]{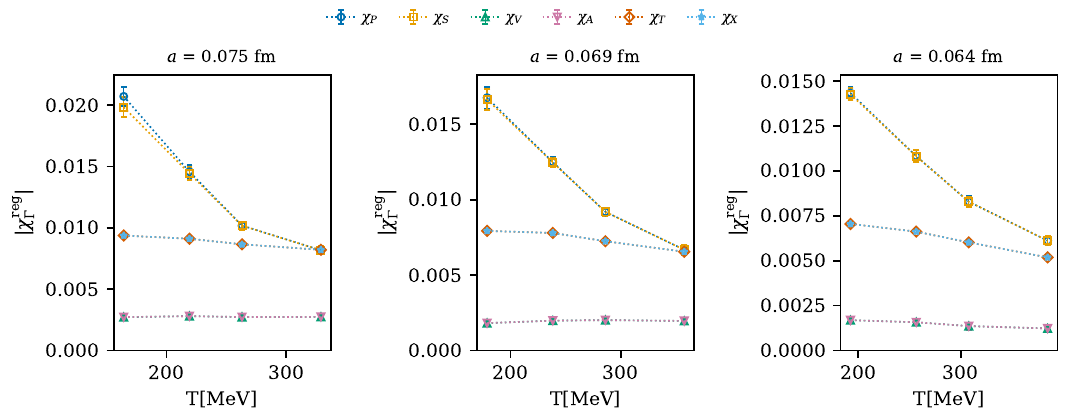}
\caption{\label{fig:chi_ud} Regularized susceptibilities of $\bar u \Gamma d$ (\ref{eq:chi_reg_lat})
    for three lattice spacings $a$=(0.075, 0.069, 0.064) fm and twelve temperatures from 164--385 MeV.
    Magnitudes are plotted, following the manifestly-positive
    normalization of ref.~\cite{Chiu:2026upk}: for $T<T_r$ the subtracted
    $\chi_\Gamma^\reg(T;T_r)$ is positive in the $(P,S)$ channels and negative in $(V,A)$ and
    $(T,X)$, and it is these magnitudes that enter $\kappa_{AB}$. The signed values are given
    in tables~\ref{tab:chi_PS_ud}--\ref{tab:chi_TX_ud}.
    The dotted lines connecting the data points in each channel are shown only to guide the eye.}
\end{figure*}

Within the resolution of our data, no splitting is observed inside the $(P,S)$, $(V,A)$, or $(T,X)$ doublets, 
except for a mild deviation in the $(P,S)$ channel at $T = 164$~MeV on the coarsest lattice 
($a = 0.075$~fm), shown in the upper panel of figure~\ref{fig:Ct_ud}. 
Since $(P,S)$ degeneracy signals $U(1)_A$ restoration, this residual $(P,S)$ splitting
at $T=164$~MeV is of direct physical interest. We note, however, that with the present
ensemble set temperature and lattice spacing are not varied independently, so the
figure alone cannot attribute the deviation to lattice spacing rather than temperature.
As discussed in section~\ref{subsec:topology} below, the dominant effect is neither: near the
crossover the difference $\chi_\pi-\chi_\delta$ is saturated by the exact zero modes of
the Dirac operator, which the valence domain-wall quarks used here reproduce only
approximately, so the $(P,S)$ splitting seen in the correlator understates the true
splitting.

To quantify the degree of degeneracy in a way that is independent of the renormalization
of the operators, a renormalization-group (RG) invariant measure is required---one that
allows quantitative comparison of symmetry breaking across different channels and across
lattice spacings. Standard probes, such as thermal hadron masses or single-time correlator values, 
can be ambiguous or sensitive to analysis details. 
Instead, we employ the RGI symmetry ratio $\kappa_{AB}$ 
defined in eq.~(\ref{eq:kappa_AB}), which integrates spectral information, provides clear normalization, 
and enables direct comparison between $SU(2)_L \times SU(2)_R$ and $U(1)_A$ breaking in all channels.

\begin{figure*}[tbp]
\centering
  \includegraphics[width=\linewidth,clip=true]{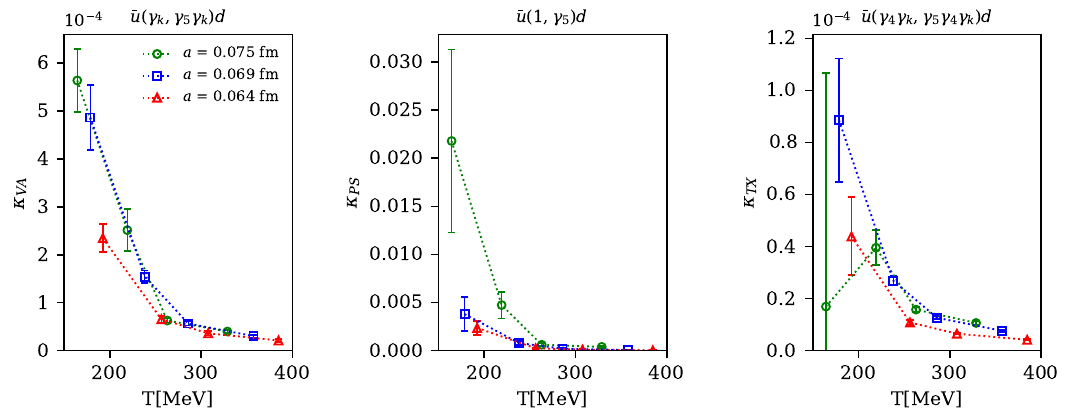}
\caption{\label{fig:kAB_ud}
RGI symmetry ratios $\kappa_{VA}$ and $\{ \kappa_{PS}, \kappa_{TX} \}$ 
for the $SU(2)_L \times SU(2)_R$ and $U(1)_A$ chiral symmetries of $(u,d)$ quarks,
computed at three lattice spacings $a$ = (0.075, 0.069, 0.064) fm and twelve temperatures in the range
164--385~MeV. Within each channel the ratio decreases toward finer lattice spacing at the
lower temperatures. Since a finer lattice spacing at fixed $N_t$ is also a higher
temperature, this trend does not separate the lattice-spacing dependence from the
temperature dependence. Note that the $\kappa_{PS}$ values are obtained with valence domain-wall
quarks, which underestimate $\chi_\pi-\chi_\delta$ near the crossover
(section~\ref{subsec:topology}). The dotted lines connecting the data points of a given
lattice spacing are shown only to guide the eye.}
\end{figure*}

Accordingly, we compute the regularized susceptibilities (\ref{eq:chi_reg_lat}) 
$\chi_P^\reg$, $\chi_S^\reg$, $\chi_V^\reg$, $\chi_A^\reg$, $\chi_T^\reg$, and $\chi_X^\reg$, shown in figure~\ref{fig:chi_ud}, 
as well as the RGI symmetry ratios $\kappa_{PS}$, $\kappa_{VA}$, and $\kappa_{TX}$ 
presented in figure~\ref{fig:kAB_ud}. These quantities are obtained for three lattice spacings 
$a=(0.075,0.069,0.064)$~fm and twelve temperatures from 164 to 385~MeV.

Numerical values of the correlators, bare susceptibilites (\ref{eq:chi_bare_lat}) 
at the reference temperature $T_r = 658$~MeV, 
regularized susceptibilities (\ref{eq:chi_reg_lat}), and RGI symmetry ratios are provided in 
appendix~\ref{app:tables}.

In figure~\ref{fig:chi_ud}, the regularized susceptibilities show no splitting within any doublet, 
again except for the $(P,S)$ channel at $T=164$~MeV on the coarsest lattice. 
This aligns with the behavior seen in the correlators and reinforces the interpretation that 
the observed $(P,S)$ deviation at the lowest temperature reflects mainly the incomplete
reproduction of the exact zero modes by the valence domain-wall quarks
(section~\ref{subsec:topology}).

\label{subsec:results_kappa}
From figure~\ref{fig:kAB_ud}, a clear hierarchy emerges:
$$
\kappa_{PS} > \kappa_{VA} > \kappa_{TX},
$$
which holds for all three lattice spacings and across the entire temperature range. 
The relative size of $\kappa_{AB}$ across channels does not by itself order the
restoration of the two symmetries. Each ratio measures the breaking in its own pair of
operators, and the pairs differ in how the breaking enters: the $(P,S)$ splitting is tied
to the topological susceptibility by an exact identity (section~\ref{subsec:topology}),
whereas the $(V,A)$ and $(T,X)$ splittings are not. A reading of the hierarchy as a
difference in restoration temperatures would, with $\kappa_{TX}$, have $U(1)_A$
effectively restored while $SU(2)_L\times SU(2)_R$ is still broken, which is not possible,
since the chiral condensate breaks $U(1)_A$ as well. The ordering is therefore not to be
interpreted as a difference in restoration temperature (section~\ref{subsec:hierarchical}).
At $T=164$~MeV, $\kappa_{TX}=1.7(9.0)\times10^{-5}$ (table~\ref{tab:chi_TX_ud}) is consistent
with zero. Its central value lies below those at $179$ and $192$~MeV by less than one standard
deviation, and these temperatures belong to different lattice spacings, so no non-monotonic
temperature dependence is implied.
At fixed $N_t$ the separation between $\kappa_{PS}$ and $\kappa_{VA}$ narrows toward
finer lattice spacing: $\kappa_{PS}/\kappa_{VA}$ is about $9.7$, $3.5$ and $2.7$ for
$a=0.075$, $0.069$ and $0.064$~fm at $N_t=10$, and $9.7$, $1.7$ and $1.7$ at $N_t=8$
(tables~\ref{tab:chi_PS_ud} and \ref{tab:chi_VA_ud}). The separation between
$\kappa_{VA}$ and $\kappa_{TX}$ does not: $\kappa_{VA}/\kappa_{TX}$ is about $3.98$,
$4.47$ and $5.54$ at $N_t=10$, and $3.75$, $3.98$ and $5.09$ at $N_t=8$
(tables~\ref{tab:chi_VA_ud} and \ref{tab:chi_TX_ud}). Since a finer lattice spacing at
fixed $N_t$ is also a higher temperature, and at each lattice spacing the ratios of the
$\kappa$'s change with $T$, neither trend can be attributed to the lattice spacing by
itself. Two contributions to the ordering can be identified with the present data:
discretization effects, and the incomplete zero-mode content of the valence domain-wall
quarks, which suppresses $\kappa_{PS}$ near the crossover
(section~\ref{subsec:topology}). Disentangling them is left to the overlap-valence
computation and to the fixed-temperature ensembles of ref.~\cite{Chiu:inprep}, as discussed
in section~\ref{subsec:future}. We
therefore present $\kappa_{PS}$, $\kappa_{VA}$, and $\kappa_{TX}$ as
finite-lattice-spacing results.

\subsection{Independence of the reference temperature}
\label{subsec:Tr_indep}

The regularized susceptibilities, and hence $\kappa_{AB}$, are defined by subtracting the
bare susceptibility at a reference temperature $T_r$, eq.~(\ref{eq:chi_reg_lat}). The
reference is a genuine choice, and it is important to establish in what sense the results
do not depend on it. The analytic answer is given in the companion
paper~\cite{Chiu:2026upk}: for any two admissible reference temperatures $T_r$ and $T_r'$,
both far enough above $T_c$ that the partners are degenerate there, the sign of the
partner splitting $\chi_A^\reg-\chi_B^\reg$, the zero $T_0$ of $\kappa_{AB}(T)$, and its
vanishing above $T_0$ are independent of the reference at each lattice spacing. Only the magnitude below $T_0$ depends on $T_r$, through
the normalization $s(T)-s(T_r)$ of the denominator. A reference is admissible if it lies
sufficiently above $T_c$ that the partner difference $d(T_r)$ is consistent with zero, and
if it retains enough time slices, $N_t=1/(aT_r)$, that the cutoff effects of the reference
susceptibilities remain small. A reference at very small $N_t$, such as $N_t=2$ where all
channels collapse to a common value, is not admissible~\cite{Chiu:2026upk}. Crucially,
these statements hold \emph{at fixed lattice spacing} and require no continuum limit. The
continuum limit enters only if one further extrapolates $\kappa_{AB}$ to $a=0$, and it is
there that the reference must be held fixed in physical units rather than at fixed $aT_r$,
since a fixed $aT_r$ sends $T_r\to\infty$ as $a\to0$ and drives $\kappa_{AB}\propto a^2\to0$
trivially at every temperature~\cite{Chiu:2026upk}. For this reason we adopt the common
physical reference $T_r=658$~MeV at all three lattice spacings (Sec.~\ref{sec:lattice},
table~\ref{tab:chi_Tr}), interpolating the reference susceptibility log-linearly in $N_t$,
between the $N_t=6$ and $N_t=4$ ensembles, where $T_r$ does not coincide with an ensemble.

We demonstrate this reference independence directly on our data, without any continuum
extrapolation, by recomputing $\kappa_{PS}$, $\kappa_{VA}$, and $\kappa_{TX}$ with three
admissible references, $T_r=658$, $715$, and $770$~MeV. These are the temperatures $1/(4a)$
of the $N_t=4$ ensembles at $\beta=6.15$, $6.18$, and $6.20$ respectively, so that each
reference is realized directly on one lattice spacing and interpolated on only the other
two, which keeps the interpolation systematic smaller than for a reference that must be
interpolated at all three. Since the reference susceptibilities enter only as constants,
this requires no new measurements at the temperatures of interest, only the bare
susceptibilities at the additional references.

The comparison is summarized in table~\ref{tab:Tr_compare}, over all twelve temperatures and
all three lattice spacings, the additional reference susceptibilities themselves being
collected in table~\ref{tab:chi_Tr_extra} of appendix~\ref{app:tables}, and
figure~\ref{fig:kAB_Tr} shows the comparison
directly. For $\kappa_{PS}$ the three references agree closely: the mean
relative difference between any pair is $4$--$10\%$ and no individual point differs by more
than $0.5\sigma$. For $\kappa_{VA}$ and $\kappa_{TX}$ the relative differences are larger,
$12$--$38\%$ in the mean, and reach $2$--$5\sigma$ on individual points. The size of the
spread follows from the structure of eq.~(\ref{eq:chi_reg_lat}). The numerator of
$\kappa_{AB}$ is unchanged by the choice of reference, since the partners are degenerate
there, and the denominator changes by $\Delta s_r=s(T_r')-s(T_r)$, the change of the bare
partner sum between the two references. The relative spread is therefore
$|\Delta s_r|/|s(T)-s(T_r)|$. For $(P,S)$ the bare sum falls steeply from $T$ to $T_r$
and is nearly flat between the references ($\Delta s_r=-2.0\times10^{-3}$ at
$a=0.064$~fm from $658$ to $770$~MeV, against $|s(T)-s(T_r)|=0.6$--$2.0\times10^{-2}$,
tables~\ref{tab:chi_Tr}, \ref{tab:chi_Tr_extra} and \ref{tab:chi_PS_ud}). For $(V,A)$
and $(T,X)$ the bare sum rises from $T$ to $T_r$, so that $\chi^\reg<0$, and continues to
rise between the references ($\Delta s_r=+0.9\times10^{-3}$ and $+2.7\times10^{-3}$),
while the regularized sums are smaller, $1.2$--$2.8\times10^{-3}$ for $(V,A)$ and
$5.2$--$9.4\times10^{-3}$ for $(T,X)$. The $(V,A)$ spread is the largest because its
denominator is the smallest, the leading free thermal coefficient of this channel being
zero, $\tilde c_0^{V,A}=0$~\cite{Chiu:2026upk}; the $(T,X)$ spread is intermediate because
its denominator is comparable to that of $(P,S)$ while $\Delta s_r$ is larger. In every channel the spread
grows with the separation of the two references, and to good accuracy in proportion to it:
the mean relative difference per MeV of separation is $0.34\%$ for $\kappa_{VA}$ and
$0.22\%$ for $\kappa_{TX}$, the same for all three pairs to within a few percent. This is
the signature of a smooth change of normalization rather than a breakdown of the
cancellation. What is common to all three references is the reference-independent
content: at every lattice spacing and every temperature the three references give the
same sign of the partner splitting. This confirms, at finite lattice spacing, the
reference independence of the sign predicted in Ref.~\cite{Chiu:2026upk}. The other two
reference-independent quantities, the zero $T_0$ and the vanishing above it, cannot be
tested with the present data, since no ratio vanishes in the simulated range: all are
resolved from zero except $\kappa_{TX}$ at $164$~MeV. The ordering
$\kappa_{PS}>\kappa_{VA}>\kappa_{TX}$ also holds for all three references, at all $36$
combinations of lattice spacing, reference, and temperature. This is an empirical
observation for the present data set and not a consequence of reference independence,
since the ordering is not among the reference-independent quantities. The magnitude of
$\kappa_{AB}$ below $T_0$ is meaningful only at a stated $T_r$, at finite lattice spacing
and in the continuum alike.

\begin{figure*}[tbp]
\centering
  \includegraphics[width=\linewidth,clip=true]{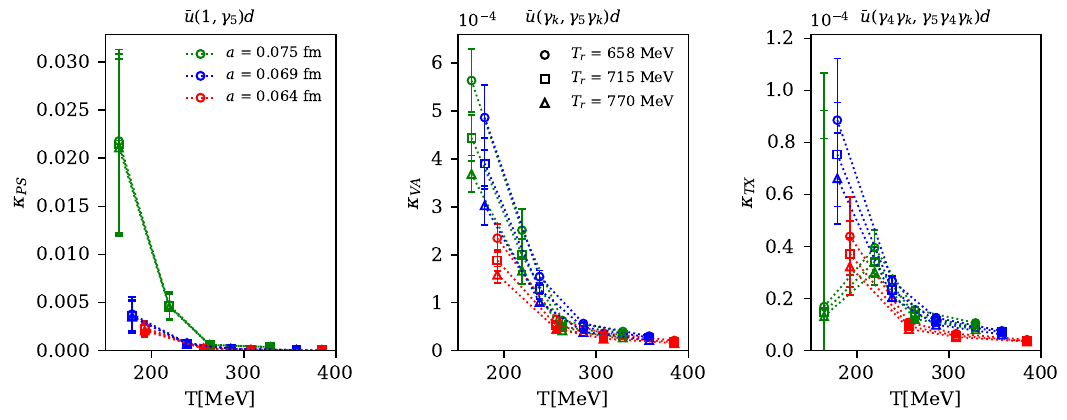}
\caption{\label{fig:kAB_Tr}
RGI symmetry ratios $\kappa_{PS}$, $\kappa_{VA}$ and $\kappa_{TX}$ at the three lattice
spacings $a=(0.075,0.069,0.064)$~fm (green, blue and red), each computed with the three
admissible reference temperatures $T_r=658$, $715$ and $770$~MeV (circles, squares and
triangles). The reference enters $\chi^\reg$ only as an additive constant, so a change of
$T_r$ rescales $\kappa_{AB}$ without altering the temperature dependence of its shape. At
every lattice spacing and every temperature the three references give the same sign of the
partner splitting and preserve the ordering $\kappa_{PS}>\kappa_{VA}>\kappa_{TX}$. They differ only in the overall
normalization, which is largest for $\kappa_{VA}$ at the lowest temperatures and which
decreases toward the effectively restored region. Table~\ref{tab:Tr_compare} quantifies the
three pairwise comparisons over all twelve temperatures. The dotted lines connecting the
data points of a given lattice spacing and reference are shown only to guide the eye.}
\end{figure*}

\begin{table*}[tbp]
\centering
\begin{adjustbox}{max width=\linewidth}
\begin{tabular}{|c|c|c|c|c|c|c|}
\hline
Channel & $T_r$ pair [MeV] & $N_{\rm pts}$ & $\langle|\Delta\kappa|/\bar\kappa\rangle$ & $\max|\Delta\kappa|/\bar\kappa$ & $\langle|\Delta\kappa|/\sigma\rangle$ & $\max|\Delta\kappa|/\sigma$ \\
\hline
$\kappa_{PS}$ & 715--770 & 12 & 3.9\% & 8.0\% & 0.10 & 0.21 \\
 & 658--715 & 12 & 5.9\% & 9.5\% & 0.15 & 0.24 \\
 & 658--770 & 12 & 9.8\% & 17.5\% & 0.25 & 0.45 \\
\hline
$\kappa_{VA}$ & 715--770 & 12 & 18.6\% & 25.1\% & 1.58 & 2.70 \\
 & 658--715 & 12 & 19.9\% & 23.8\% & 1.46 & 2.53 \\
 & 658--770 & 12 & 38.1\% & 46.4\% & 2.82 & 4.42 \\
\hline
$\kappa_{TX}$ & 715--770 & 12 & 11.7\% & 13.7\% & 1.28 & 2.99 \\
 & 658--715 & 12 & 13.3\% & 16.7\% & 1.30 & 2.60 \\
 & 658--770 & 12 & 24.9\% & 30.3\% & 2.50 & 4.60 \\
\hline
\end{tabular}
\end{adjustbox}
\caption{Reference-temperature independence at fixed lattice spacing. For each channel the RGI
ratio is recomputed with the three admissible references $T_r=658$, $715$, and $770$~MeV, and the
three pairwise comparisons are summarized over all twelve temperatures and all three lattice
spacings ($N_{\rm pts}=12$, i.e.\ $N_t=16,12,10,8$ at each spacing). $\Delta\kappa$ is the
difference between the two references and $\bar\kappa$ their mean. The $\sigma$ columns combine
the two jackknife errors in quadrature and are therefore conservative, since the two ratios share
the statistics of their numerator. The relative-difference columns are free of this. The larger
spread of $\kappa_{VA}$ and $\kappa_{TX}$ reflects the ratio $|\Delta s_r|/|s(T)-s(T_r)|$ of
the change of the bare partner sum between the two references to the regularized sum, which is
largest for $(V,A)$, whose denominator is the smallest since its leading free thermal
coefficient vanishes, $\tilde c_0^{V,A}=0$~\cite{Chiu:2026upk}, and intermediate for $(T,X)$,
whose bare sum still rises between the references (see text).}
\label{tab:Tr_compare}
\end{table*}

\section{Discussion and Outlook}
\label{sec:discussion}

In this work, we have introduced a renormalization-group invariant observable, 
the RGI symmetry ratio $\kappa_{AB}$, designed to provide a quantitative 
and scheme-independent measure of symmetry breaking in QCD.
Using this ratio, we have performed a systematic lattice study of 
the relative strength of $SU(2)_L \times SU(2)_R$ chiral symmetry breaking and $U(1)_A$ 
axial symmetry breaking in finite-temperature QCD with $N_f = 2+1+1$ flavors, 
employing optimal domain-wall fermions at the physical point.
Our analysis spans three lattice spacings and twelve temperatures in the range $164$--$385$ MeV,
covering the chiral crossover region.

We examined three independent symmetry-breaking channels 
in the \emph{nonsinglet sector with quark-connected correlators}, namely:
the scalar--pseudoscalar channel sensitive to $U(1)_A$ breaking ($\kappa_{PS}$), 
the vector--axial-vector channel probing chiral symmetry breaking ($\kappa_{VA}$), 
and an additional $U(1)_A$-sensitive tensor vector--axial-tensor vector channel ($\kappa_{TX}$).
At each lattice spacing, these channels exhibit a clear hierarchy:
$\kappa_{PS} > \kappa_{VA} > \kappa_{TX}$, which holds over the full temperature range.
Because the ensembles are generated at fixed $\beta$ with
$N_t$ varied, no two share a common physical temperature at different lattice spacings, so that a
direct fixed-temperature continuum extrapolation is not possible with the present data. We
therefore present these ratios as finite-lattice-spacing results.

Two contributions to the ordering are identified in this paper, discretization effects and the
incomplete zero-mode content of the valence domain-wall quarks, and the present ensemble set
cannot separate either from the accompanying change in temperature. Disentangling them is left
to the overlap-valence computation and the fixed-temperature ensembles of
ref.~\cite{Chiu:inprep}, as discussed in section~\ref{subsec:future}. Two caveats apply to $\kappa_{PS}$ in
particular. First, as discussed in section~\ref{subsec:topology} below, the valence domain-wall quarks used
here reproduce the exact zero modes of the Dirac operator only approximately, so the connected
difference $\chi_\pi-\chi_\delta$ extracted from them is suppressed near the crossover and
$\kappa_{PS}$ is correspondingly understated. Second, that same difference is tied by the identity
$\chi_\pi-\chi_\delta=\chi_{\rm disc}=\chi_t/m_q^2$ to the topological susceptibility, as
the preliminary results of ref.~\cite{Chiu:inprep} with the overlap low modes
confirm at $T=164$~MeV. Nonsinglet $U(1)_A$ restoration in the
scalar--pseudoscalar channel and the suppression of topological fluctuations are therefore one and
the same phenomenon, and neither is achieved at the lowest temperature simulated here.
We accordingly refrain from assigning a restoration temperature to either symmetry from the
present data, and we make no claim of a parametrically separated restoration scale for $U(1)_A$
relative to $SU(2)_L \times SU(2)_R$, in either direction.

It is worth stressing what $\kappa_{AB}$ provides that a comparison of 
screening masses does not. As an integral of the 
correlator, $\kappa_{AB}$ weighs the entire spectrum in each channel, so 
$\kappa_{AB} = 0$ is the condition that the two channels have identical 
integrated spectral weights, not merely degenerate ground states. Degeneracy 
of screening masses is a necessary but not a sufficient condition for this 
equality. Away from single-pole dominance a self-normalized shape comparison 
can even vanish while the channels remain inequivalent, since the 
normalization discards precisely the relative spectral weights that 
$\kappa_{AB}$ retains~\cite{Chiu:2026upk}. The screening-mass splitting 
and $\kappa_{AB}$ therefore answer different questions, and for the 
integrated symmetry statement made here $\kappa_{AB}$ is the appropriate 
observable, with a controlled continuum limit and exact RG 
invariance~\cite{Chiu:2026upk}.

Within a renormalization-group framework, a degeneracy of $\kappa_{PS}$, $\kappa_{VA}$, and
$\kappa_{TX}$ would imply that $SU(2)_L \times SU(2)_R$ and $U(1)_A$ symmetry-breaking effects
for \emph{quark-connected correlators} become comparably suppressed near $T_c$, and that the
corresponding symmetry-breaking operators acquire similar infrared relevance for nonsinglet
observables. Establishing such a degeneracy requires the continuum limit that the present
ensemble set does not permit.

\subsection{Connections to topological susceptibility and singlet sector}
\label{subsec:topology}

An important subtlety involves the quark-disconnected parts of the scalar singlet ($\sigma$) 
and pseudoscalar singlet ($\eta$) mesons in $N_f=2$ QCD.
Recall that the physical $\eta'$ corresponds to the $\eta$ singlet in $N_f=2$ QCD.
Throughout this subsection we use the prescription that sums over all time
slices, eq.~(\ref{eq:chi_bare_lat_all}), including the $t=0$ contact term. 
This is the natural choice here, because the index-theorem
relation~\eqref{eq:Qt_Trg5} involves the full lattice trace
$\mathrm{Tr}(\cdots)\equiv\sum_x\mathrm{tr}(\cdots)$ over all sites, including
coincident points, and retaining the $t=0$ slice keeps both sides of the
identities in the same convention. With the contact term excluded the
disconnected sum would no longer reproduce the topological susceptibility
$\chi_t$, since the $t=0$ contribution, which includes its share of the zero-mode
term, would be missing. We therefore adopt the all-slice
prescription~(\ref{eq:chi_bare_lat_all}) for the identities below. The physical
conclusions hold for either prescription, as noted at the end of this subsection.

Upon effective restoration of \emph{full} $SU(2)_L \times SU(2)_R$ chiral symmetry,
eqs.~(\ref{eq:chi_sigma_pi})--(\ref{eq:chi_rho_a1}) hold. Those involving both nonsinglet
and singlet channels, eqs.~(\ref{eq:chi_sigma_pi}) and (\ref{eq:chi_eta_delta}), yield
\begin{align}
\chi_\sigma - \chi_{\delta} = \chi_\pi - \chi_\delta = \chi_\pi - \chi_{\eta},
\label{eq:chi_sigma_delta_pi_eta_SU2}
\end{align}
which relates the quark-disconnected parts of $\chi_\sigma$ and $\chi_{\eta}$:
\begin{align}
\chi_{\text{disc}} \equiv \chi_\sigma - \chi_{\delta} = \chi_\pi - \chi_{\eta} \equiv \chi_{5,\text{disc}},
\label{eq:chi_chi5_disc}
\end{align}
where
\begin{align}
\chi_{\text{disc}} &= \frac{1}{V} \left\{ \left\langle [\mathrm{Tr} (D_c + m_q)^{-1}]^{2} \right\rangle
                   - \left\langle \mathrm{Tr} (D_c + m_q)^{-1} \right\rangle^2 \right\}, \label{eq:chidisc} \\
\chi_{5,\text{disc}} &= \frac{1}{V} \left\{\left\langle[\mathrm{Tr}\gamma_5 (D_c + m_q)^{-1}]^{2}\right\rangle
                 - \left\langle \mathrm{Tr} \gamma_5 (D_c + m_q)^{-1} \right\rangle^2 \right\}.
\label{eq:chi5disc}
\end{align}

In lattice QCD with exact chiral symmetry, the topological charge $Q_t$ satisfies
\begin{align}
Q_t = m_q \mathrm{Tr} \gamma_5 (D_c + m_q)^{-1}, \quad \forall m_q > 0.
\label{eq:Qt_Trg5}
\end{align}
Assuming eq.~(\ref{eq:Qt_Trg5}) can be measured precisely (e.g., with overlap fermions via the index theorem), 
eq.~(\ref{eq:chi_chi5_disc}) implies
\begin{align}
\chi_{\text{disc}} = \chi_{5,\text{disc}} = \frac{\chi_t}{m_q^2},
\label{eq:chi_chi5_disc_chit}
\end{align}
where $\chi_t$ is the topological susceptibility, 
\begin{align}
\chi_t = \frac{1}{V} \left( \langle Q_t^2 \rangle - \langle Q_t \rangle^2 \right).
\label{eq:chi_t}
\end{align}
These relations are written for the bare susceptibilities. The temperature
subtraction~(\ref{eq:chi_reg}) removes the additive divergences of $\chi_\pi$ and
$\chi_\delta$, but it also subtracts the topological contribution $\chi_t(T_r)/m_q^2$ to the
reference splitting $\chi_\pi(T_r)-\chi_\delta(T_r)$. With eq.~(\ref{eq:chi_R}), $Z_P=Z_S$
and $m_q=Z_S^{-1}m_q^R$, which follows from the RG invariance of $m_q\bar qq$, one obtains
\begin{equation}
\big[\chi_\pi^R(T)-\chi_\delta^R(T)\big]-\big[\chi_\pi^R(T_r)-\chi_\delta^R(T_r)\big]
=\frac{\chi_t(T)-\chi_t(T_r)}{(m_q^R)^2},
\label{eq:pidelta_reg_Tr}
\end{equation}
with the renormalization scale suppressed (cf.\ ref.~\cite{Chiu:2026upk}). The renormalized
relation $\chi_\pi^R-\chi_\delta^R=\chi_t/(m_q^R)^2$ is therefore recovered only if
$\chi_t(T_r)\ll\chi_t(T)$, so that $\chi_\pi(T_r)$ cancels $\chi_\delta(T_r)$. Otherwise the
regularized splitting measures $\chi_t(T)-\chi_t(T_r)$ rather than $\chi_t(T)$. This
requires $T_r\gg T_c$, the condition already imposed on the reference temperature in
section~\ref{subsec:integrated}. At $T_r=658$~MeV, $\chi_P$ and $\chi_S$ agree within
uncertainties at every lattice spacing (table~\ref{tab:chi_Tr}).

Thus, upon effective restoration of \emph{full} $SU(2)_L \times SU(2)_R$ chiral symmetry, 
$\chi_t/m_q^2$ equals any of the three differences: 
$\chi_\sigma - \chi_{\delta}$, $\chi_\pi - \chi_\delta$, or $\chi_\pi - \chi_{\eta}$.
In particular, the connected, nonsinglet difference $\chi_\pi-\chi_\delta$ that enters
$\kappa_{PS}$ is equal to the quark-disconnected $\chi_{\rm disc}=\chi_{5,{\rm disc}}$.
The two degeneracies on which this rests, $\chi_\pi=\chi_\sigma$ and
$\chi_\delta=\chi_\eta$, are the susceptibility form of the equalities among two-point
correlators that follow from the classification of interpolating currents into chiral
multiplets, derived systematically by Cohen and Ji~\cite{CohenJi:1997}.
The relation $\chi_\pi-\chi_\delta=\chi_{\rm disc}=\chi_{5,{\rm disc}}$ in this susceptibility 
form was stated and used by the HotQCD collaboration with
domain-wall fermions~\cite{Bazavov:2012qja}, and the equality
$\chi_\pi-\chi_\delta=\chi_{\rm disc}$ has been confirmed numerically after continuum and
chiral extrapolation with HISQ fermions~\cite{Ding:2020xlj}.

The further step to the topological susceptibility, $\chi_\pi-\chi_\delta=\chi_t/m_q^2$,
appears as eq.~(43) of ref.~\cite{Bazavov:2012qja}, obtained there by combining the
$SU(2)_L\times SU(2)_R$ relation with $\chi_t=m_l^2\chi_{5,{\rm disc}}$, which in the
continuum follows from integrating the anomalous axial Ward identity and on the lattice holds
exactly only for an operator obeying the index theorem, eq.~(\ref{eq:Qt_Trg5}). It is not
confirmed numerically in that work. On the
contrary, the $\chi_t$ measured there from smeared gauge links exceeds
$m_l^2\chi_{5,{\rm disc}}$ by more than a factor of two near the crossover, and the authors
conclude (their appendix~D) that their $\chi_t$ is contaminated by unphysical short-distance
fluctuations at $T\lesssim T_c$, uncorrelated with the fermionic degrees of freedom, while
the fermionic susceptibilities agree among themselves at the $10$--$20\%$ level. What has
been verified numerically is therefore
$\chi_\pi-\chi_\delta=\chi_{\rm disc}=\chi_{5,{\rm disc}}$, not the relation to $\chi_t$
itself. The same is true of ref.~\cite{Ding:2020xlj}, which works with highly improved
staggered quarks---an action with neither exact zero modes nor an index theorem---and
tests $\chi_\pi-\chi_\delta=\chi_{\rm disc}$ against the measured Dirac eigenvalue
density without measuring $\chi_t$.

Ref.~\cite{Buchoff:2013nra} reports the same pattern with the same domain-wall action:
$\chi_t$ and $\tilde m_l^2\chi_{5,{\rm disc}}$ disagree substantially at $T>0$, with the
fractional discrepancy decreasing toward higher temperature and finer lattice spacing.
The relation to $\chi_t$ has therefore, to our knowledge, not yet been verified at nonzero
temperature.

A direct test of this step therefore requires $Q_t$ from the index theorem, with both sides of
eq.~(\ref{eq:chi_chi5_disc_chit}) evaluated from the same Dirac spectrum on the same
configurations, so that every unit of topological charge is carried by an exact zero mode
and the ambiguity diagnosed in ref.~\cite{Bazavov:2012qja} does not arise. The lowest
eigenmodes of the overlap Dirac operator provide exactly this. Their computation on the
present ensembles is the subject of work in preparation~\cite{Chiu:inprep}. Here we
set out the relations it tests and their consequences for $\kappa_{PS}$, and quote its
preliminary results where they bear on the interpretation of our data.

At first sight eq.~(\ref{eq:chi_sigma_delta_pi_eta_SU2}) presents a puzzle: how can
$\chi_\pi-\chi_\delta$, built only from quark-connected correlators, equal
$\chi_{\rm disc}=\chi_t/m_q^2$, which arises solely from quark-disconnected diagrams?
The resolution is that both are saturated by the same exact zero modes of the Dirac
operator. For a chirally symmetric valence operator $D_c$ with $n_\pm$ zero modes of
definite chirality, the connected difference has the spectral representation
\begin{align}
\chi_\pi-\chi_\delta=\frac{1}{V}\sum_{\lambda}\frac{m_q^2}{(\lambda^2+m_q^2)^2}
=\frac{|Q_t|}{V m_q^2}+\frac{1}{V}\sum_{\lambda\neq 0}\frac{m_q^2}{(\lambda^2+m_q^2)^2},
\label{eq:pidelta_spectral}
\end{align}
where the sum runs over all eigenmodes and $|Q_t|=|n_+-n_-|$ is the number of exact zero
modes. The zero-mode term carries a $1/m_q^2$ pole, and the disconnected
$\chi_{5,{\rm disc}}=\chi_t/m_q^2$, with $\chi_t$ of eq.~(\ref{eq:chi_t}), carries the same
pole. The two agree whenever the nonzero-mode remainder in eq.~(\ref{eq:pidelta_spectral}) is
negligible and $\langle|Q_t|\rangle\simeq V\chi_t$, i.e.\ in the dilute
topological regime above the crossover. The identity is therefore not a statement that
connected and disconnected diagrams coincide diagram by diagram, but that near and above
$T_c$ both are dominated by the same topological zero modes.

Equating eq.~(\ref{eq:pidelta_spectral}) with $\chi_t/m_q^2$, the identity requires
\begin{align}
\frac{\chi_\pi-\chi_\delta}{\chi_t/m_q^2}
=\frac{\langle|Q_t|\rangle}{V\chi_t}+\frac{m_q^2\,\chi_{\rm nz}}{\chi_t}=1,
\qquad
\chi_{\rm nz}\equiv\frac{1}{V}\sum_{\lambda\neq0}\frac{m_q^2}{(\lambda^2+m_q^2)^2},
\label{eq:ward_ratio}
\end{align}
which combines two physically distinct conditions. The nonzero-mode term $\chi_{\rm nz}$ is
suppressed as $(m_q/\lambda)^4$ for every mode with $\lambda\gg m_q$ and vanishes as a spectral
gap opens, i.e.\ with $SU(2)_L\times SU(2)_R$ restoration in the light sector. The zero-mode
term is the dilute-topological condition: if every configuration carries $|Q_t|\le1$, then
$\langle|Q_t|\rangle=\langle Q_t^2\rangle\ge V\chi_t$, with equality up to the
parity-violating fluctuation $\langle Q_t\rangle^2$, and the ratio falls below unity only
through the sectors with $|Q_t|\ge2$. In a dilute instanton gas, where $Q_t$ follows a Skellam
distribution, $\langle|Q_t|\rangle/(V\chi_t)$ is fixed by $V\chi_t$ alone and rises toward
unity as topological fluctuations are suppressed. At fixed temperature, however,
$\langle|Q_t|\rangle\propto\sqrt{V}$ while $V\chi_t\propto V$, so the zero-mode term falls
with the volume, and whether the ratio approaches unity in the infinite-volume limit depends
on whether the near-zero modes of instanton--anti-instanton pairs grow to compensate. That
they should is supported by ref.~\cite{Ding:2020xlj}, which finds a peaked structure in
$\rho(\lambda\to0,m_l)$ with amplitude proportional to $m_l^2$, sharpening as $a\to0$, i.e.\
the $\rho\sim m_l^2\delta(\lambda)$ of a weakly interacting instanton gas, which is the
continuum counterpart of the exact zero modes.

Equation~(\ref{eq:ward_ratio}) therefore gives the conditions for $U(1)_A$ restoration in
the $(\pi,\delta)$ channel in two steps. First, $\chi_{\rm nz}$ vanishes as the spectral gap
opens, which is $SU(2)_L\times SU(2)_R$ restoration in the light sector. Second, the zero-mode
term saturates $\chi_t/m_q^2$, which on the present volumes holds because every configuration
carries $|Q_t|\le1$, while on larger volumes the near-zero modes of
instanton--anti-instanton pairs must take over. Once both hold,
$\chi_\pi-\chi_\delta=\chi_t/m_q^2$, and $U(1)_A$ is restored in this channel exactly when
$\chi_t$ vanishes. Since $\chi_t$ decreases with temperature but does not vanish at any
finite $T$, this restoration is effective, in the sense of section~\ref{subsec:chi_sym}, and
never exact. At finite lattice spacing $\chi_t$ is in addition enhanced by discretization
effects, which hold up the nonsinglet $U(1)_A$ breaking beyond its continuum value. For these
reasons the ratio~(\ref{eq:ward_ratio}) itself is not a practical probe of
$SU(2)_L\times SU(2)_R$ restoration: its temperature dependence is dominated by the dilution
of topology rather than by the symmetry, since the zero-mode term is fixed by $V\chi_t$, the
symmetry-sensitive term $m_q^2\chi_{\rm nz}/\chi_t$ is divided by the same shrinking
$\chi_t$, and the configurations with $|Q_t|\ge2$, which carry the departure from unity,
become rare. We therefore probe $SU(2)_L\times SU(2)_R$ restoration with $\kappa_{VA}$, and
use the $(P,S)$ and $(T,X)$ channels for $U(1)_A$.

The zero-mode saturation of $\chi_\pi-\chi_\delta$ has a direct consequence for
$\kappa_{PS}$. The valence domain-wall quarks used to
compute $\chi_\pi$ and $\chi_\delta$ in section~\ref{sec:results} satisfy the
Ginsparg--Wilson relation only approximately at finite $N_s$, and consequently do not
reproduce the exact zero modes of $D_c$. The $1/m_q^2$ pole in
eq.~(\ref{eq:pidelta_spectral}) is largely absent from the valence correlators. Near the
crossover, where $\chi_t/m_q^2$ is sizeable, the connected difference $\chi_\pi-\chi_\delta$
obtained with valence domain-wall quarks therefore \emph{underestimates} the true value,
and the $\kappa_{PS}$ values obtained from the regularized susceptibilities of
table~\ref{tab:chi_PS_ud} (and figure~\ref{fig:kAB_ud}) are correspondingly suppressed at the
lowest temperatures. The preliminary results of ref.~\cite{Chiu:inprep} support this
picture. With the overlap low modes on the $T=164$~MeV ensemble, $\chi_\pi-\chi_\delta$ is
saturated by the exact zero modes, changing by less than $0.01\%$ between the lowest $5$ and
$100$ conjugate pairs, and it equals $\chi_t/m_q^2$ to within $14\%$, the deficit being
accounted for by the configurations with $|Q_t|\ge2$. On the $T=179$~MeV ensemble, where
every configuration carries $|Q_t|\le1$, the two agree within $3\%$. The valence
domain-wall difference on the $T=164$~MeV ensemble, summed over the plateau at $t\ge1$,
reproduces only about $0.3\%$ of the zero-mode-saturated value. This, rather than a
discretization effect, is the dominant origin of the mild $(P,S)$ deviation noted in
section~\ref{sec:results}.
For the same reason, a small $\kappa_{PS}$ from valence domain-wall quarks is not by
itself evidence of $U(1)_A$ restoration, and it should not be compared directly with
determinations that use overlap valence quarks, such as ref.~\cite{Aoki:2020noz}. That
comparison awaits the valence overlap computation~\cite{Chiu:inprep}. Each exact zero
mode contributes $+1/(2Vm_q^2)$ to $\chi_P$ and $-1/(2Vm_q^2)$ to $\chi_S$. Where they
outweigh the nonzero-mode part of $\chi_S^\reg$, it turns negative while
$\chi_P^\reg>0$, and $\kappa_{PS}=1$ by eq.~(\ref{eq:kappa_AB}).

The consequence for the interpretation of $\kappa_{PS}$ is the central point of this
subsection. Because $\chi_\pi-\chi_\delta=\chi_t/m_q^2$, the nonsinglet
$U(1)_A$-breaking difference cannot vanish while $\chi_t$ remains finite: the two are
tied together by the identity and vanish at the \emph{same} temperature. A small
valence domain-wall $\kappa_{PS}$ at the lowest simulated temperature therefore reflects
the incomplete zero-mode content of the valence operator, not the restoration of
$U(1)_A$. Correspondingly, we do not find support for a two-stage scenario in which the nonsinglet $(\pi,\delta)$ degeneracy
precedes the vanishing of $\chi_t$. For the scalar--pseudoscalar pair, nonsinglet
$U(1)_A$ restoration and the suppression of topological fluctuations are one and the same
phenomenon. We emphasize that it is the relation
$\chi_\pi-\chi_\delta=\chi_t/m_q^2$, with $|Q_t|$ counted by the index theorem, that
removes the two-stage possibility in this channel. The fermionic chain
$\chi_\pi-\chi_\delta=\chi_{\rm disc}=\chi_{5,{\rm disc}}$ alone cannot: it relates
quantities that are all built from the light-quark propagator and says nothing about
$\chi_t$, and in a formulation without exact zero modes both sides can be reproduced from
the near-zero spectral density without reference to the index theorem.

The vector and tensor pairs couple to the exact zero modes differently from $(\pi,\delta)$
and from each other, so the zero-mode content of the valence operator may affect
$\kappa_{VA}$ and $\kappa_{TX}$ as well. Settling this requires valence quarks that carry the
exact zero modes in every channel, and it will be addressed with valence overlap quarks
preconditioned by the low modes~\cite{Chiu:inprep}. The values of $\kappa_{VA}$ and
$\kappa_{TX}$ in section~\ref{sec:results} are therefore presented as obtained with the
valence domain-wall quarks, without a zero-mode correction, and the ordering of the three
ratios is to be read with this in mind.

Although the identities above were written for the all-slice
prescription~(\ref{eq:chi_bare_lat_all}), which is the natural one for the
topological susceptibility, the conclusion that $\chi_\pi-\chi_\delta$ is tied to
$\chi_t$ does not depend on this choice: the zero-mode contribution to the
connected difference is a property of the Dirac spectrum, independent of how the
bare susceptibility is summed. The two prescriptions differ only in
short-distance bookkeeping that cancels from the physical conclusions.

\subsection{Restoration pattern in the nonsinglet sector}
\label{subsec:hierarchical}

The finite-lattice-spacing results for $\kappa_{PS}$, $\kappa_{VA}$, and $\kappa_{TX}$
(section~\ref{sec:results}) show all three ratios decreasing with temperature above the
crossover, with the ordering $\kappa_{PS}>\kappa_{VA}>\kappa_{TX}$ at each lattice spacing.
The vector--axial-vector ratio $\kappa_{VA}$ probes $SU(2)_L\times SU(2)_R$, while
$\kappa_{PS}$ and $\kappa_{TX}$ probe $U(1)_A$ through two different Dirac structures.
As discussed in section~\ref{subsec:topology}, $\chi_\pi-\chi_\delta$ carries the full
$1/m_q^2$ zero-mode pole and is tied to $\chi_t$ by eq.~(\ref{eq:chi_chi5_disc_chit}), and
$\kappa_{PS}$ from the valence domain-wall quarks is suppressed near the crossover because
they do not reproduce the exact zero modes. How the zero modes enter the vector and tensor
channels, and whether the observed ordering of the three ratios survives when the valence
quarks carry them, will be settled by the valence overlap
computation~\cite{Chiu:inprep}. We therefore do not interpret the ordering in terms
of a difference in restoration temperature between the symmetries.

For $N_f=2+1+1$ QCD at the physical point the chiral transition is a smooth crossover, so
none of the differences $\chi_\sigma-\chi_\delta$, $\chi_\pi-\chi_\delta$, or
$\chi_\pi-\chi_\eta$ vanishes identically. Each diminishes gradually as $T$ increases
above $T_c$. Lattice determinations of $\chi_t(T)$ are affected by significant
discretization effects and a consistent continuum picture has not yet emerged across
studies (see e.g.\ \cite{Petreczky:2016vrs,Borsanyi:2016ksw,Chen:2022fid,Athenodorou:2022aay,Kotov:2025ilm}),
but for $N_f=2+1$ and $N_f=2+1+1$ QCD at physical masses $\chi_t(T)$ remains sizeable up to
$T\gg T_c$, falling off with a power law broadly consistent with the dilute instanton gas
prediction~\cite{Gross:1980br}. By eq.~(\ref{eq:chi_chi5_disc_chit}) the same gradual
decrease then governs the nonsinglet $(\pi,\delta)$ splitting: $\chi_\pi-\chi_\delta$
tracks $\chi_t/m_q^2$ and does not vanish before it, as the preliminary low-mode results
of ref.~\cite{Chiu:inprep} confirm at the lowest simulated temperature.

Whether the $SU(2)_L\times SU(2)_R$ ratio $\kappa_{VA}$ and the $U(1)_A$ ratio
$\kappa_{TX}$, which are not tied to $\chi_t$ by an identity like
eq.~(\ref{eq:chi_chi5_disc_chit}), restore at the same temperature as
$\kappa_{PS}$ or at a different one is a question the present finite-lattice-spacing
data cannot settle, because the observed ordering of the three ratios is not separated
from discretization effects and from the incomplete zero-mode content of the valence
domain-wall quarks (sections~\ref{sec:results} and \ref{subsec:topology}). We therefore refrain from
assigning a restoration temperature to either symmetry from these data.

What can be said about $\chi_t$ above $T_c$ follows from an exact identity. In lattice QCD
with exact chiral symmetry, the topological susceptibility obeys
(appendix A of ref.~\cite{Mao:2009sy})
\begin{align}
2N_f\,\chi_{\eta'}=\sum_f\frac{\Sigma_f}{m_f}-\Big(\sum_f\frac{1}{m_f}\Big)^2\chi_t ,
\label{eq:chit_LS_2111}
\end{align}
where $\chi_{\eta'}$ is the susceptibility of the flavor-singlet pseudoscalar
$\bar q\gamma_5 q$ of all $N_f$ flavors, with $t^0$ implied as in
section~\ref{subsec:correlators}, and $\Sigma_f=\langle\mathrm{Tr}(D_c+m_f)^{-1}\rangle/V$.
In ref.~\cite{Mao:2009sy} the singlet is normalized as
$\eta_0=N_f^{-1}\sum_f\bar q_f\gamma_5q_f$, whose zero-momentum correlator is
$G_{\eta'}(0)=(2/N_f)\chi_{\eta'}$. If the $\eta'$ stays massive, $\chi_{\eta'}$ has no pole
in the light-quark mass. In the chirally broken phase the terms singular in the light-quark
mass must then cancel, which gives the generalization of the Leutwyler--Smilga
relation~\cite{Leutwyler:1992yt} to nondegenerate flavors,
$\chi_t=(\sum_f\Sigma_f/m_f)(\sum_f1/m_f)^{-2}$, with leading term $\chi_t\simeq m_l\Sigma_l/2$
for two light flavors of mass $m_l$. The heavy-flavor terms enter this relation at relative
order $m_l/m_s$ and $m_l/m_c$, the same order as the $\eta'$ term that it leaves undetermined.
Above the light-flavor crossover, $T>T_c^{ud}$, with $T_c^{ud}$ the temperature denoted $T_c$
elsewhere in this paper, the identity does not determine $\chi_t$ by itself. The
quark-connected part of $2N_f\chi_{\eta'}$ equals $\sum_f\Sigma_f/m_f$ exactly, so
eq.~(\ref{eq:chit_LS_2111}) identifies $(\sum_f1/m_f)^2\chi_t$ with the quark-disconnected
part of $2N_f\chi_{\eta'}$. It shows, however, that $\chi_t$, and with it
$\chi_\pi-\chi_\delta=\chi_t/m_l^2$, remains nonzero as long as $\chi_{\eta'}$ differs from
its connected part, i.e., as long as $U(1)_A$ breaking persists in the flavor-singlet
channel. Since the $\eta'$ contains the strange and charm quarks, whose chiral symmetries are
restored only at higher temperatures, this is expected to hold in the window
$T_c^{ud}<T<T_c^{udsc}$, where $T_c^{udsc}$ is the temperature at which the chiral symmetry
of all four flavors is effectively restored~\cite{Chiu:2026upk}.

The strange and charm quarks enter $\chi_\pi-\chi_\delta$ only through the sea. The
susceptibilities $\chi_\pi$ and $\chi_\delta$ are built from the valence $u$ and $d$ quarks,
whereas $\chi_t$ is a property of the gauge-field ensemble, which is generated with the
fermion determinants of all four flavors. In a dilute instanton gas, for example, each
instanton carries one zero mode per flavor, and each flavor that is light compared with the
inverse instanton size contributes a factor of its mass to the instanton
weight~\cite{Gross:1980br}. Hence $\chi_t\propto m_l^2$, and $\chi_\pi-\chi_\delta=\chi_t/m_l^2$
stays finite as $m_l\to0$, with a magnitude set by the instanton density, to which the
strange and charm quarks contribute through the sea. How large this contribution is remains
a question for direct measurement, which the overlap low-mode study of
ref.~\cite{Chiu:inprep} can address.

\paragraph{Probing the singlet sector}
The three ratios studied here are built from quark-connected, nonsinglet correlators.
Full effective restoration of $SU(2)_L\times SU(2)_R$ and $U(1)_A$ involves in addition
six channels with singlet operators. Four probe $SU(2)_L\times SU(2)_R$ between singlet
and nonsinglet partners, $\kappa_{\sigma,\pi}$, $\kappa_{\eta,\delta}$,
$\kappa_{\omega_T,b_1}$ and $\kappa_{h_{1T},\rho_T}$, and two probe $U(1)_A$ in the singlet
sector, $\kappa_{\sigma,\eta}$ and $\kappa_{\omega_T,h_{1T}}$. All six are RG-invariant
(section~\ref{subsec:renorm}), and all require quark-disconnected diagrams, the dominant
computational challenge. The channels involving the scalar singlet $\sigma$ also require
VEV subtraction. By eq.~(\ref{eq:chi_chi5_disc_chit}) the mixed singlet--nonsinglet
differences $\chi_\sigma-\chi_\delta$ and $\chi_\pi-\chi_\eta$ equal $\chi_\pi-\chi_\delta$
once $SU(2)_L\times SU(2)_R$ is effectively restored, so they carry no information beyond
the nonsinglet difference and $\chi_t$. The genuinely new information is in the singlet
$U(1)_A$ ratios $\kappa_{\sigma,\eta}$ and $\kappa_{\omega_T,h_{1T}}$, whose measurement we
leave to future work.

\subsection{Comparison with other symmetry patterns above $T_c$}
\label{subsec:broader}

Above the crossover, lattice studies have identified further symmetry patterns that are
not symmetries of the QCD Lagrangian. Approximate chiral-spin $SU(2)_{\mathrm{CS}}$
symmetry~\cite{Glozman:2014mka}, observed in $N_f=2$~\cite{Rohrhofer:2019qwq,Rohrhofer:2019qal},
$N_f=2+1+1$~\cite{Chiu:2023hnm,Chiu:2024jyz} and
$N_f=2+1+1+1$~\cite{Chiu:2024bqx} lattice QCD, emerges only in an intermediate window,
typically $T\sim2T_c$ to $T\sim4T_c$ for light quarks, where chromoelectric binding
dominates, and it fades once thermal energies exceed that binding. The infrared symmetric
phase reported around $T_{\mathrm{IR}}\sim230$~MeV in $N_f=2+1$ lattice
QCD~\cite{Alexandru:2019gdm,Meng:2023nxf} points to the same window of correlated,
quasi-hadronic physics. These are distinct from the chiral symmetries studied here, whose
breaking decreases monotonically with temperature rather than appearing and disappearing.

Topological studies are complementary. Center-vortex analyses~\cite{tHooft:1977nqb} in $N_f=2+1$ QCD find a
pronounced change in vortex percolation at $\gtrsim2T_c$, with a sparse network surviving
to higher temperatures~\cite{Mickley:2024vkm}, so topological fluctuations remain
non-negligible well into the deconfined regime. This is consistent with the picture of
section~\ref{subsec:topology}: the nonsinglet $(\pi,\delta)$ splitting tracks $\chi_t$ and
therefore persists as long as topology does.

\subsection{Work in preparation and future extensions}
\label{subsec:future}

The present paper is confined to the valence optimal domain-wall quarks. Two studies that
build on it are in preparation~\cite{Chiu:inprep}. They are left out here because they require the overlap low
modes, or the overlap propagators, on every ensemble of table~\ref{tab:ensembles}, and they
will be presented when these are complete.
\begin{itemize}
  \item \textbf{Overlap low modes and the $(\pi,\delta)$ splitting:}
   The lowest eigenmodes of the overlap Dirac operator on all ensembles give the connected
   difference $\chi_\pi-\chi_\delta$ of the $(u,d)$ sector from the spectral
   representation~(\ref{eq:pidelta_spectral}), and $Q_t$ from the index theorem on the same
   configurations. This tests $\chi_\pi-\chi_\delta=\chi_t/m_q^2$, eq.~(\ref{eq:ward_ratio}),
   at every temperature and lattice spacing, including its dependence on the sectors with
   $|Q_t|\ge2$ and on the volume. For the light quarks the nonzero modes enter with the weight
   $(m_q/\lambda)^4$, so the difference is saturated by the lowest ${\cal O}(100)$ conjugate
   pairs. Its preliminary results are quoted in section~\ref{subsec:topology}.
  \item \textbf{Valence overlap quarks with low-mode preconditioning:}
   Computing the nonsinglet susceptibilities in all six channels with valence overlap quarks,
   using the projected low modes as a preconditioner, so that the exact zero modes enter every
   channel at every temperature and lattice spacing. This removes the valence systematic of
   section~\ref{subsec:topology} from $\kappa_{PS}$, determines how the zero modes enter the
   vector and tensor channels and hence $\kappa_{VA}$ and $\kappa_{TX}$, and settles whether
   the ordering of the three ratios survives. It also supplies $\chi_t$ from the index theorem
   on the same configurations at no extra cost.
\end{itemize}
Beyond these, two extensions are needed:
\begin{itemize}
  \item \textbf{Ensembles for a fixed-$T$ continuum limit:} Ensembles at several fixed
   physical temperatures, with the bare coupling tuned at each $N_t$ so that three or more
   lattice spacings are available at each temperature, and larger spatial volumes
   ($40^3$ and $64^3$). This is the standard setting for a direct continuum extrapolation
   of $\kappa_{PS}$, $\kappa_{VA}$ and $\kappa_{TX}$ at fixed $T$, which the present
   fixed-$\beta$ ensembles do not permit. The larger volumes also control the
   finite-volume dependence of the zero-mode and near-zero-mode contributions discussed in
   section~\ref{subsec:topology}.
  \item \textbf{Singlet channels:} With the previous steps in place, the six
   singlet-involving ratios ($\kappa_{\sigma,\pi}$, $\kappa_{\eta,\delta}$,
   $\kappa_{\omega_T,b_1}$, $\kappa_{h_{1T},\rho_T}$, $\kappa_{\sigma,\eta}$,
   $\kappa_{\omega_T,h_{1T}}$) can be computed from quark-disconnected diagrams with
   stochastic all-to-all propagators and colour--Dirac dilution, again with overlap
   quarks preconditioned by the low modes, completing the restoration pattern.
\end{itemize}

\subsection{RG interpretation and comparison with Pisarski--Wilczek analyses}
\label{subsec:rg_pw}

The renormalization-group analysis of Pisarski and Wilczek (PW)~\cite{Pisarski:1983ms}
shows that the order of the two-flavor chiral transition is controlled by the RG scaling of
the $U(1)_A$-breaking operator at the $O(4)$ fixed point. If that operator is
\emph{relevant}, the infrared physics is governed by the stable $O(4)$ Wilson--Fisher fixed
point and a second-order transition in the $O(4)$ universality class is allowed; if it is
\emph{irrelevant}, the effective symmetry enlarges to $U(2)_L\times U(2)_R$, for which the
$\epsilon$-expansion finds no infrared-stable fixed point and hence a fluctuation-induced
first-order transition. Pelissetto and Vicari~\cite{Pelissetto:2013hqa} established that for
$N_f=2$ the operator is \emph{marginally irrelevant} at the $O(4)$ fixed point, implying
approximate $O(4)$ scaling in the chiral limit near the PW critical temperature.

This framework applies strictly in the chiral limit at the infrared fixed point, and it
classifies universality rather than observables at finite quark mass. In physical QCD the
transition is a crossover and no critical fixed point exists, so restoration must be
defined operationally. Here it is the vanishing of $\kappa_{AB}$ within statistical
uncertainties (section~\ref{subsec:integrated}). That condition is substantially weaker
than RG irrelevance: it implies neither an enlargement of the symmetry to
$U(2)_L\times U(2)_R$ nor a change of universality class, only that the breaking is not
resolved by the present observables and precision. The PW/PV analysis also draws no
distinction between quark-connected and quark-disconnected contributions, which is the
distinction that organizes our results.

With that understood, the present finite-lattice-spacing results speak to operational
restoration in the nonsinglet sector only, and in this respect they stand on the same footing
as the single-lattice-spacing determinations of refs.~\cite{Aoki:2020noz,Chiu:2023hnm}. The temperature at which each ratio becomes
unresolvable is a crossover scale, not a critical temperature, and we assign no common
value to it: the ordering of the three ratios is not separated from discretization effects
and from the incomplete zero-mode content of the valence domain-wall quarks
(section~\ref{sec:results}), and $\kappa_{PS}$ obtained with valence domain-wall quarks
understates the $(\pi,\delta)$ splitting near the crossover, where that splitting equals
$\chi_t/m_q^2$. The anomaly is therefore not confined to the singlet sector: it enters the
nonsinglet sector through the exact zero modes. Complementary scaling studies are
consistent with this reading, finding approximate $O(4)$ scaling at physical quark masses
in $N_f=2+1+1$ QCD~\cite{Kotov:2021rah}, a continuum $N_f=2$ chiral-limit transition
compatible with $O(N)$ scaling~\cite{Klinger:2025xxb}, and, in the functional RG,
axial-anomaly effects persisting throughout a broad crossover region~\cite{Braun:2020ada}.

Full operational restoration, including the singlet sector, requires
$\kappa_{\sigma,\eta}$, $\kappa_{\omega_T,h_{1T}}$, $\kappa_{\sigma,\pi}$,
$\kappa_{\eta,\delta}$, $\kappa_{\omega_T,b_1}$ and $\kappa_{h_{1T},\rho_T}$ to vanish
together with $\chi_t$, all of which involve quark-disconnected diagrams and the anomaly
directly. Existing studies find topological fluctuations persisting up to
$T\gtrsim2T_c$~\cite{Petreczky:2016vrs,Borsanyi:2016ksw,Chen:2022fid,Athenodorou:2022aay,Kotov:2025ilm},
but a continuum determination combining chirally symmetric fermions with overlap-based
topology in both sea and valence sectors is still lacking. Whether the eventual
restoration of the singlet channels bears any relation to the $U(2)_L\times U(2)_R$
scenario of the chiral-limit analysis remains open.

\subsection{Concluding remarks}

The RGI symmetry ratio $\kappa_{AB}$ introduced here provides a flexible and systematically 
improvable framework for quantifying symmetry breaking in QCD. 
Applied to finite-temperature QCD, it leads to several central conclusions:

\begin{enumerate}
\item \textbf{RG-invariant diagnostics:}
The $\kappa_{AB}$ ratios furnish renormalization-group invariant, scheme-independent measures of 
symmetry breaking across different operator channels, allowing direct comparison between lattice spacings 
and formulations.

\item \textbf{Finite-lattice-spacing hierarchy:}
At each of the three lattice spacings the nonsinglet ratios obey
$\kappa_{PS}>\kappa_{VA}>\kappa_{TX}$ over the full temperature range. Because the ensembles are generated at fixed
$\beta$ with $N_t$ varied, no two share a common physical temperature at different lattice
spacings, and a direct fixed-temperature continuum extrapolation is deferred to a
dedicated ensemble set with $\beta$ tuned to common target temperatures.

\item \textbf{The $(\pi,\delta)$ splitting and the topological susceptibility:}
Effective $SU(2)_L\times SU(2)_R$ restoration implies
$\chi_\pi-\chi_\delta=\chi_{\rm disc}=\chi_{5,{\rm disc}}$, which follows from the chiral
classification of interpolating currents~\cite{CohenJi:1997} and has been confirmed
numerically with domain-wall~\cite{Bazavov:2012qja} and HISQ~\cite{Ding:2020xlj} fermions.
Through the index theorem it gives $\chi_\pi-\chi_\delta=\chi_t/m_q^2$, and both sides
of this relation are saturated by the exact zero modes of the Dirac operator. It is this
relation, rather than the purely fermionic chain, that ties the nonsinglet splitting to
topology: the nonsinglet $U(1)_A$-breaking difference vanishes together with $\chi_t$, not
before it. The valence domain-wall quarks, which reproduce the exact zero modes only
approximately, understate this splitting near the crossover, and $\kappa_{PS}$ computed from
them is correspondingly suppressed. Preliminary results with the overlap low
modes~\cite{Chiu:inprep}, which count $|Q_t|$ by the index theorem rather than from
gauge-link smearing, confirm the relation at $T=164$ and $179$~MeV. The singlet ratios
$\kappa_{\sigma,\eta}$ and $\kappa_{\omega_T,h_{1T}}$ remain to be measured.
\end{enumerate}

The results presented here are those of the valence optimal domain-wall quarks. What has
been left out will be presented in follow-up work now in preparation~\cite{Chiu:inprep}
(section~\ref{subsec:future}). The first uses the overlap low modes on all ensembles to study
the $(P,S)$ channel of the $(u,d)$ sector and to test $\chi_\pi-\chi_\delta=\chi_t/m_q^2$
directly. The second uses overlap valence quarks with low-mode
preconditioning to determine $\kappa_{AB}$ in all channels, including the role of the exact
zero modes in the vector and tensor channels. A dedicated ensemble
set, with $\beta$ tuned to common target temperatures and larger volumes, will then permit a
fixed-temperature continuum extrapolation, and the quark-disconnected diagrams needed for the
singlet ratios can be computed with stochastic all-to-all propagators using the same
preconditioned overlap quarks. Carried through in that order, the $\kappa_{AB}$ framework
offers a systematic strategy for mapping the restoration pattern of chiral and axial
symmetries in QCD.

\begin{acknowledgments}
We are grateful to Academia Sinica Grid Computing Center
and National Center for High Performance Computing for the computer time and facilities.
This work was supported by the National Science and Technology Council
(Grants No.~108-2112-M-003-005, No.~109-2112-M-003-006, No.~110-2112-M-003-009),
and Academia Sinica Grid Computing Centre (Grant No.~AS-CFII-112-103).
\end{acknowledgments}

\vfill
\clearpage

\appendix

\section{Meson Operator Notation}
\label{app:notation}
\begin{itemize}
\item For $N_f = 2$ flavor, the generators are
$t^0 = \Id/2$ (for flavor singlet) and
$t^a = \tau^a/2$ ($a = 1,2,3$ for flavor nonsinglets), where $\Id$ is the $2\times2$ identity matrix
and $\tau^a$ are the Pauli matrices. They satisfy the normalization condition
$\tr(t^a t^b) = \delta^{ab}/2$ for all $a,b = 0,1,2,3$.
\item The flavor-singlet generator $t^0$ is always suppressed in our notation for singlet operators; 
e.g., we write $\bar{q} q$ to mean $\bar{q} t^0 q$, and similarly $\bar{q} \gamma_5 q$ for $\bar{q} \gamma_5 t^0 q$.
\item Spatial index $i=1,2,3$ (or $k=1,2,3$) is implicit for the vector-type channels ($V$, $A$, $T$, $X$).
\item Singlet operators ($\sigma$, $\eta$, $\omega$, $h_1$, $\omega_T$, $h_{1T}$) involve 
      quark-disconnected diagrams. The $\sigma$ requires VEV subtraction, and the disconnected
      part of the $\eta$ is related to $\chi_t$ (section~\ref{subsec:topology}).
\item Nonsinglet operators involve only quark-connected diagrams.
\item The tensor-vector representations $\rho_T^a$ (via $T_i$) and $b_1^a$ (via $X_i$) probe $U(1)_A$ through 
      a different Dirac structure than the scalar--pseudoscalar pair ($\delta$--$\pi$).
\item The vector singlet $\omega$ and axial-vector singlet $h_1$ are \emph{not} $U(1)_A$ partners,
      since both currents are invariant under $U(1)_A$ ($\{\gamma_5, \gamma_\mu\} = 0$).
      In contrast, the tensor vector singlet $\omega_T$ and axial-tensor vector singlet $h_{1T}$ \emph{are}
      $U(1)_A$ partners ($\{\gamma_5, \gamma_4\gamma_k\} \neq 0$), analogous to 
      the scalar--pseudoscalar pair in the tensor vector channel.
\end{itemize}

\begin{table*}[tbp]
\centering
\caption{Summary of meson operators and their corresponding susceptibilities in $N_f=2$ QCD. 
The notation follows eqs.~(\ref{eq:chi_sigma_pi})--(\ref{eq:chi_omega_h1}) in the main text. 
Singlet ($s$) and nonsinglet ($ns$) channels are distinguished, with the latter involving only quark-connected diagrams. 
Tensor operators $T_k$ and $X_k$ ($k=1,2,3$) are defined with spatial index $k$.
The interpolating operators are listed up to a conventional Hermiticity phase:
the $\gamma_5$-containing densities ($P$, $A$, $X$) acquire a factor $i$ when
written as Hermitian operators~\cite{Chiu:2026upk}. This phase cancels in the
susceptibilities $\chi_\Gamma\propto\langle O_\Gamma O_\Gamma^\dagger\rangle$ and does not
affect any result below.}
\label{tab:meson_notation}
\begin{adjustbox}{max width=\linewidth}
\begin{tabular}{lllll}
\toprule
\textbf{Meson} & \textbf{Operator} & \textbf{Channel} & \textbf{Flavor} & \textbf{Susceptibility} \\
\midrule
$\pi^a$ & $\bar{q}\gamma_5 t^a q$ & Pseudoscalar ($P$) & Nonsinglet ($ns$) & $\chi_\pi = \chi_P^{ns}$ \\
$\delta^a $ & $\bar{q} t^a q$ & Scalar ($S$) & Nonsinglet ($ns$) & $\chi_\delta = \chi_S^{ns}$ \\
$\sigma$ & $\bar{q}q$ & Scalar ($S$) & Singlet ($s$) & $\chi_\sigma = \chi_S^{s}$ \\
$\eta$ ($N_f=2$) & $\bar{q}\gamma_5 q$ & Pseudoscalar ($P$) & Singlet ($s$) & $\chi_\eta = \chi_P^{s}$ \\
\midrule
$\rho^a$ & $\bar{q}\gamma_i t^a q$ & Vector ($V_i$) & Nonsinglet ($ns$) & $\chi_\rho = \chi_{V_i}^{ns}$ \\
$a_1^a$ & $\bar{q}\gamma_5\gamma_i t^a q$ & Axial-vector ($A_i$) & Nonsinglet ($ns$) & $\chi_{a_1} = \chi_{A_i}^{ns}$ \\
$\omega$ & $\bar{q}\gamma_i q$ & Vector ($V_i$) & Singlet ($s$) & $\chi_\omega = \chi_{V_i}^{s}$ \\
$h_1$ & $\bar{q}\gamma_5\gamma_i q$ & Axial-vector ($A_i$) & Singlet ($s$) & $\chi_{h_1} = \chi_{A_i}^{s}$ \\
\midrule
$\rho_T^a$ & $\bar{q}\gamma_4\gamma_i t^a q$ & Tensor vector ($T_i$) & Nonsinglet ($ns$) & $\chi_{\rho_T} = \chi_{T_i}^{ns}$ \\
$b_1^a$ & $\bar{q}\gamma_5\gamma_4\gamma_i t^a q$ & Axial-tensor vector ($X_i$) & Nonsinglet ($ns$) & $\chi_{b_1} = \chi_{X_i}^{ns}$ \\
$\omega_T$ & $\bar{q}\gamma_4\gamma_i q$ & Tensor vector ($T_i$) & Singlet ($s$) & $\chi_{\omega_T} = \chi_{T_i}^{s}$ \\
$h_{1T}$ & $\bar{q}\gamma_5\gamma_4\gamma_i q$ & Axial-tensor vector ($X_i$) & Singlet ($s$) & $\chi_{h_{1T}} = \chi_{X_i}^{s}$ \\
\bottomrule
\end{tabular}
\end{adjustbox}
\end{table*}

\vfill
\clearpage

\section{Tables of results}
\label{app:tables}

In this appendix, we provide numerical data of the figures in the main text. 
Specifically, the data for the time-correlation functions, 
bare susceptibilities at the reference temperature, 
regularized susceptibilities, 
and RGI symmetry ratios for $u,d$ off-diagonal flavor-nonsinglet mesons 
are presented in tabular form. 

The numerical values for the time-correlation functions $C_{\Gamma}(t)$ of the $\bar u \Gamma d$ 
on the $32^3 \times 16$ lattice, shown in figure \ref{fig:Ct_ud}, are tabulated in 
tables \ref{tab:Ct_ud_b615}--\ref{tab:Ct_ud_b620} for the $(P, S, V, A, T, X)$ channels,  
for three lattice spacings $a$ = (0.075, 0.069, 0.064) fm. All correlators and susceptibilities tabulated here
are those of the operators $O^a_\Gamma=\bar q\,\Gamma\,t^a q$ of
eq.~(\ref{eq:Ct}), whose flavor generators satisfy $\tr(t^at^b)=\delta^{ab}/2$, and
therefore carry the factor $1/2$ relative to the unnormalized bilinear
$\bar u\Gamma d$. The ratios $\kappa_{AB}$ are unaffected by this normalization, which
cancels between numerator and denominator.

Numerical results for the bare susceptibilities (\ref{eq:chi_bare_lat}) 
$(\chi_P, \chi_S, \chi_V, \chi_A, \chi_T, \chi_X)$ 
at the common physical reference temperature $T_r = 658$~MeV are listed in
table~\ref{tab:chi_Tr}, with the two members of each pair shown separately, and those at
the two higher reference temperatures $T_r'=715$ and $770$~MeV, used for the comparison of
section~\ref{subsec:Tr_indep}, in table~\ref{tab:chi_Tr_extra}.

Numerical results for the regularized susceptibilities (\ref{eq:chi_reg_lat}) 
$(\chi_S^\reg, \chi_P^\reg, \chi_V^\reg, \chi_A^\reg, \chi_T^\reg, \chi_X^\reg)$, 
presented in figure \ref{fig:chi_ud}, are listed in tables \ref{tab:chi_PS_ud}--\ref{tab:chi_TX_ud}. 
These tables also include the corresponding RGI symmetry ratios $\kappa_{PS}$, $\kappa_{VA}$, 
and $\kappa_{TX}$, shown in figure~\ref{fig:kAB_ud}. 
Data are provided for the same three lattice spacings across twelve temperatures ranging from 164 to 385 MeV.

Statistical uncertainties are estimated using the jackknife method with a bin size of 
$5$--$15$ configurations of which the statistical error saturates.
In tables~\ref{tab:chi_PS_ud}--\ref{tab:chi_TX_ud}, the susceptibilities 
$\chi_A^\reg$ and $\chi_B^\reg$ are quoted to eight significant figures, more than 
strictly warranted by their individual jackknife errors. 
This choice is intentional and necessary for reproducibility: the symmetry 
ratio $\kappa_{AB}$ of eq.~(\ref{eq:kappa_AB}) is formed from the 
\emph{difference} of two nearly-equal susceptibilities, and this difference 
becomes extremely small as the symmetry is restored at high temperature. 
For example, at $T = 385$~MeV the relative splitting 
$|\chi_A^\reg - \chi_B^\reg|/|\chi_A^\reg|$ is of order $10^{-5}$ in the PS channel and 
smaller still in the VA and TX channels. 
Quoting $\chi_A^\reg$ and $\chi_B^\reg$ to only five or six significant figures would 
therefore round away the very difference that determines $\kappa_{AB}$, so 
that the tabulated $\kappa_{AB}$ could not be reconstructed from the 
tabulated $\chi_A^\reg$ and $\chi_B^\reg$. 
With eight significant figures, the reader can reproduce each $\kappa_{AB}$ 
from the corresponding $\chi_A^\reg$ and $\chi_B^\reg$ to well within its jackknife 
uncertainty, across all channels and temperatures. 
The same precision also makes manifest the subtle differences between 
symmetry partners (such as $\chi_V^\reg$ and $\chi_A^\reg$) that would otherwise be 
obscured within the quoted error bars.

\begin{table*}[tbp]
\setlength{\tabcolsep}{3pt}
\centering
\footnotesize  % This will apply to the table content
\begin{adjustbox}{max width=\linewidth}
\begin{tabular}{|ccccccc|}
\hline
$t/a$ & $C_P(t)$ & $C_S(t)$  & $C_V(t)$ & $C_A(t)$  & $C_T(t)$ & $C_X(t)$  \\
\hline 
  1 & $1.963(7) \times 10^{-2}$ & $1.960(7) \times 10^{-2}$ & $1.4643(44) \times 10^{-2}$ & $1.4643(44) \times 10^{-2}$ & $1.0969(21) \times 10^{-2}$ & $1.0969(21) \times 10^{-2}$ \\
  2 & $3.231(46) \times 10^{-3}$ & $3.199(47) \times 10^{-3}$ & $1.645(12) \times 10^{-3}$ & $1.645(12) \times 10^{-3}$ & $9.855(70) \times 10^{-4}$ & $9.855(70) \times 10^{-4}$ \\
  3 & $1.306(43) \times 10^{-3}$ & $1.276(42) \times 10^{-3}$ & $4.159(38) \times 10^{-4}$ & $4.157(38) \times 10^{-4}$ & $2.193(25) \times 10^{-4}$ & $2.193(25) \times 10^{-4}$ \\
  4 & $8.100(420) \times 10^{-4}$ & $7.815(415) \times 10^{-4}$ & $1.649(18) \times 10^{-4}$ & $1.649(18) \times 10^{-4}$ & $8.410(130) \times 10^{-5}$ & $8.410(130) \times 10^{-5}$ \\
  5 & $6.190(415) \times 10^{-4}$ & $5.920(410) \times 10^{-4}$ & $8.770(115) \times 10^{-5}$ & $8.760(115) \times 10^{-5}$ & $4.58(10) \times 10^{-5}$ & $4.58(10) \times 10^{-5}$ \\
  6 & $5.310(410) \times 10^{-4}$ & $5.060(405) \times 10^{-4}$ & $5.800(85) \times 10^{-5}$ & $5.795(85) \times 10^{-5}$ & $3.17(8) \times 10^{-5}$ & $3.17(8) \times 10^{-5}$ \\
  7 & $4.91(41) \times 10^{-4}$ & $4.67(41) \times 10^{-4}$ & $4.59(8) \times 10^{-5}$ & $4.59(8) \times 10^{-5}$ & $2.62(7) \times 10^{-5}$ & $2.62(8) \times 10^{-5}$ \\
  8 & $4.80(40) \times 10^{-4}$ & $4.57(41) \times 10^{-4}$ & $4.26(8) \times 10^{-5}$ & $4.25(8) \times 10^{-5}$ & $2.46(7) \times 10^{-5}$ & $2.46(7) \times 10^{-5}$ \\
  9 & $4.95(40) \times 10^{-4}$ & $4.71(40) \times 10^{-4}$ & $4.62(9) \times 10^{-5}$ & $4.62(9) \times 10^{-5}$ & $2.62(9) \times 10^{-5}$ & $2.62(9) \times 10^{-5}$ \\
  10 & $5.385(385) \times 10^{-4}$ & $5.130(395) \times 10^{-4}$ & $5.880(90) \times 10^{-5}$ & $5.875(90) \times 10^{-5}$ & $3.17(10) \times 10^{-5}$ & $3.17(10) \times 10^{-5}$ \\
  11 & $6.295(375) \times 10^{-4}$ & $6.020(390) \times 10^{-4}$ & $8.875(105) \times 10^{-5}$ & $8.870(105) \times 10^{-5}$ & $4.57(10) \times 10^{-5}$ & $4.57(10) \times 10^{-5}$ \\
  12 & $8.225(360) \times 10^{-4}$ & $7.925(375) \times 10^{-4}$ & $1.663(16) \times 10^{-4}$ & $1.662(16) \times 10^{-4}$ & $8.405(125) \times 10^{-5}$ & $8.405(125) \times 10^{-5}$ \\
  13 & $1.323(36) \times 10^{-3}$ & $1.290(38) \times 10^{-3}$ & $4.192(27) \times 10^{-4}$ & $4.191(27) \times 10^{-4}$ & $2.212(21) \times 10^{-4}$ & $2.212(21) \times 10^{-4}$ \\
  14 & $3.260(37) \times 10^{-3}$ & $3.225(40) \times 10^{-3}$ & $1.661(9) \times 10^{-3}$ & $1.661(9) \times 10^{-3}$ & $9.980(50) \times 10^{-4}$ & $9.980(55) \times 10^{-4}$ \\
  15 & $1.970(6) \times 10^{-2}$ & $1.966(5) \times 10^{-2}$ & $1.4698(36) \times 10^{-2}$ & $1.4697(36) \times 10^{-2}$ & $1.1010(22) \times 10^{-2}$ & $1.1010(22) \times 10^{-2}$ \\
\hline
\end{tabular}
\end{adjustbox}
\caption{Time-correlation function $C_\Gamma(t)$~(\ref{eq:Ct}) of the off-diagonal
nonsinglet operators $\bar u \Gamma t^a d$ 
for $P,S,V,A,T,X$ channels, including the factor $1/2$ from the flavor normalization
$\tr(t^at^b)=\delta^{ab}/2$ of eq.~(\ref{eq:Ct}), on the $32^3 \times 16$ lattice with lattice spacing $a$=0.075 fm.}
\label{tab:Ct_ud_b615}
\end{table*}

\begin{table*}[tbp]
\setlength{\tabcolsep}{3pt}
\centering
\footnotesize  % This will apply to the table content
\begin{adjustbox}{max width=\linewidth}
\begin{tabular}{|ccccccc|}
\hline
$t/a$ & $C_P(t)$ & $C_S(t)$  & $C_V(t)$ & $C_A(t)$  & $C_T(t)$ & $C_X(t)$  \\
\hline 
  1 & $1.972(5) \times 10^{-2}$ & $1.972(5) \times 10^{-2}$ & $1.4801(36) \times 10^{-2}$ & $1.4801(36) \times 10^{-2}$ & $1.1042(22) \times 10^{-2}$ & $1.1042(22) \times 10^{-2}$ \\
  2 & $3.152(32) \times 10^{-3}$ & $3.147(32) \times 10^{-3}$ & $1.688(8) \times 10^{-3}$ & $1.688(8) \times 10^{-3}$ & $1.0030(47) \times 10^{-3}$ & $1.0029(47) \times 10^{-3}$ \\
  3 & $1.184(30) \times 10^{-3}$ & $1.180(29) \times 10^{-3}$ & $4.319(29) \times 10^{-4}$ & $4.319(29) \times 10^{-4}$ & $2.240(19) \times 10^{-4}$ & $2.239(19) \times 10^{-4}$ \\
  4 & $6.745(275) \times 10^{-4}$ & $6.710(270) \times 10^{-4}$ & $1.732(13) \times 10^{-4}$ & $1.731(13) \times 10^{-4}$ & $8.595(95) \times 10^{-5}$ & $8.590(95) \times 10^{-5}$ \\
  5 & $4.80(26) \times 10^{-4}$ & $4.77(26) \times 10^{-4}$ & $9.250(80) \times 10^{-5}$ & $9.245(80) \times 10^{-5}$ & $4.64(7) \times 10^{-5}$ & $4.63(7) \times 10^{-5}$ \\
  6 & $3.92(25) \times 10^{-4}$ & $3.88(25) \times 10^{-4}$ & $6.110(65) \times 10^{-5}$ & $6.105(65) \times 10^{-5}$ & $3.17(6) \times 10^{-5}$ & $3.17(6) \times 10^{-5}$ \\
  7 & $3.50(25) \times 10^{-4}$ & $3.48(25) \times 10^{-4}$ & $4.82(6) \times 10^{-5}$ & $4.81(6) \times 10^{-5}$ & $2.57(8) \times 10^{-5}$ & $2.57(8) \times 10^{-5}$ \\
  8 & $3.38(25) \times 10^{-4}$ & $3.35(25) \times 10^{-4}$ & $4.46(8) \times 10^{-5}$ & $4.46(8) \times 10^{-5}$ & $2.40(8) \times 10^{-5}$ & $2.40(8) \times 10^{-5}$ \\
  9 & $3.51(26) \times 10^{-4}$ & $3.48(25) \times 10^{-4}$ & $4.86(10) \times 10^{-5}$ & $4.85(10) \times 10^{-5}$ & $2.58(9) \times 10^{-5}$ & $2.58(9) \times 10^{-5}$ \\
  10 & $3.94(26) \times 10^{-4}$ & $3.90(26) \times 10^{-4}$ & $6.195(115) \times 10^{-5}$ & $6.195(115) \times 10^{-5}$ & $3.20(9) \times 10^{-5}$ & $3.20(9) \times 10^{-5}$ \\
  11 & $4.86(27) \times 10^{-4}$ & $4.80(27) \times 10^{-4}$ & $9.385(155) \times 10^{-5}$ & $9.380(155) \times 10^{-5}$ & $4.71(7) \times 10^{-5}$ & $4.71(7) \times 10^{-5}$ \\
  12 & $6.815(290) \times 10^{-4}$ & $6.760(285) \times 10^{-4}$ & $1.744(21) \times 10^{-4}$ & $1.744(21) \times 10^{-4}$ & $8.690(110) \times 10^{-5}$ & $8.690(110) \times 10^{-5}$ \\
  13 & $1.188(33) \times 10^{-3}$ & $1.183(32) \times 10^{-3}$ & $4.316(34) \times 10^{-4}$ & $4.315(34) \times 10^{-4}$ & $2.248(20) \times 10^{-4}$ & $2.247(20) \times 10^{-4}$ \\
  14 & $3.14(6) \times 10^{-3}$ & $3.13(5) \times 10^{-3}$ & $1.678(10) \times 10^{-3}$ & $1.678(10) \times 10^{-3}$ & $9.990(55) \times 10^{-4}$ & $9.990(55) \times 10^{-4}$ \\
  15 & $1.964(5) \times 10^{-2}$ & $1.963(5) \times 10^{-2}$ & $1.4739(37) \times 10^{-2}$ & $1.4739(37) \times 10^{-2}$ & $1.1009(23) \times 10^{-2}$ & $1.1009(23) \times 10^{-2}$ \\
\hline
\end{tabular}
\end{adjustbox}
\caption{Time-correlation function $C_\Gamma(t)$~(\ref{eq:Ct}) of the off-diagonal
nonsinglet operators $\bar u \Gamma t^a d$ 
for $P,S,V,A,T,X$ channels, including the factor $1/2$ from the flavor normalization
$\tr(t^at^b)=\delta^{ab}/2$ of eq.~(\ref{eq:Ct}), on the $32^3 \times 16$ lattice with lattice spacing $a$=0.069 fm.}
\label{tab:Ct_ud_b618}
\end{table*}

\begin{table*}[tbp]
\setlength{\tabcolsep}{3pt}
\centering
\footnotesize  % This will apply to the table content
\begin{adjustbox}{max width=\linewidth}
\begin{tabular}{|ccccccc|}
\hline
$t/a$ & $C_P(t)$ & $C_S(t)$  & $C_V(t)$ & $C_A(t)$  & $C_T(t)$ & $C_X(t)$  \\
\hline 
  1 & $1.9609(49) \times 10^{-2}$ & $1.9607(49) \times 10^{-2}$ & $1.4740(32) \times 10^{-2}$ & $1.4740(32) \times 10^{-2}$ & $1.0983(20) \times 10^{-2}$ & $1.0983(20) \times 10^{-2}$ \\
  2 & $3.119(16) \times 10^{-3}$ & $3.117(16) \times 10^{-3}$ & $1.694(8) \times 10^{-3}$ & $1.694(8) \times 10^{-3}$ & $1.0026(43) \times 10^{-3}$ & $1.0026(43) \times 10^{-3}$ \\
  3 & $1.153(16) \times 10^{-3}$ & $1.152(16) \times 10^{-3}$ & $4.367(23) \times 10^{-4}$ & $4.366(23) \times 10^{-4}$ & $2.253(14) \times 10^{-4}$ & $2.253(14) \times 10^{-4}$ \\
  4 & $6.435(135) \times 10^{-4}$ & $6.420(135) \times 10^{-4}$ & $1.754(11) \times 10^{-4}$ & $1.754(11) \times 10^{-4}$ & $8.675(80) \times 10^{-5}$ & $8.675(80) \times 10^{-5}$ \\
  5 & $4.48(13) \times 10^{-4}$ & $4.46(13) \times 10^{-4}$ & $9.305(65) \times 10^{-5}$ & $9.305(65) \times 10^{-5}$ & $4.67(9) \times 10^{-5}$ & $4.67(9) \times 10^{-5}$ \\
  6 & $3.59(13) \times 10^{-4}$ & $3.57(13) \times 10^{-4}$ & $6.050(55) \times 10^{-5}$ & $6.050(55) \times 10^{-5}$ & $3.17(9) \times 10^{-5}$ & $3.17(9) \times 10^{-5}$ \\
  7 & $3.19(13) \times 10^{-4}$ & $3.17(13) \times 10^{-4}$ & $4.68(6) \times 10^{-5}$ & $4.68(6) \times 10^{-5}$ & $2.56(8) \times 10^{-5}$ & $2.55(8) \times 10^{-5}$ \\
  8 & $3.08(13) \times 10^{-4}$ & $3.06(13) \times 10^{-4}$ & $4.28(6) \times 10^{-5}$ & $4.28(6) \times 10^{-5}$ & $2.37(11) \times 10^{-5}$ & $2.37(11) \times 10^{-5}$ \\
  9 & $3.23(14) \times 10^{-4}$ & $3.21(14) \times 10^{-4}$ & $4.62(8) \times 10^{-5}$ & $4.62(8) \times 10^{-5}$ & $2.52(11) \times 10^{-5}$ & $2.52(11) \times 10^{-5}$ \\
  10 & $3.66(14) \times 10^{-4}$ & $3.65(14) \times 10^{-4}$ & $5.930(80) \times 10^{-5}$ & $5.930(85) \times 10^{-5}$ & $3.11(11) \times 10^{-5}$ & $3.11(11) \times 10^{-5}$ \\
  11 & $4.58(15) \times 10^{-4}$ & $4.56(14) \times 10^{-4}$ & $9.125(95) \times 10^{-5}$ & $9.125(95) \times 10^{-5}$ & $4.59(13) \times 10^{-5}$ & $4.59(13) \times 10^{-5}$ \\
  12 & $6.535(145) \times 10^{-4}$ & $6.520(145) \times 10^{-4}$ & $1.732(11) \times 10^{-4}$ & $1.732(11) \times 10^{-4}$ & $8.600(120) \times 10^{-5}$ & $8.600(120) \times 10^{-5}$ \\
  13 & $1.161(15) \times 10^{-3}$ & $1.159(15) \times 10^{-3}$ & $4.346(23) \times 10^{-4}$ & $4.346(23) \times 10^{-4}$ & $2.254(18) \times 10^{-4}$ & $2.253(18) \times 10^{-4}$ \\
  14 & $3.128(15) \times 10^{-3}$ & $3.127(15) \times 10^{-3}$ & $1.698(7) \times 10^{-3}$ & $1.698(7) \times 10^{-3}$ & $1.0071(40) \times 10^{-3}$ & $1.0071(40) \times 10^{-3}$ \\
  15 & $1.9646(41) \times 10^{-2}$ & $1.9644(41) \times 10^{-2}$ & $1.4769(27) \times 10^{-2}$ & $1.4769(27) \times 10^{-2}$ & $1.1006(17) \times 10^{-2}$ & $1.1006(17) \times 10^{-2}$ \\
\hline
\end{tabular}
\end{adjustbox}
\caption{Time-correlation function $C_\Gamma(t)$~(\ref{eq:Ct}) of the off-diagonal
nonsinglet operators $\bar u \Gamma t^a d$ 
for $P,S,V,A,T,X$ channels, including the factor $1/2$ from the flavor normalization
$\tr(t^at^b)=\delta^{ab}/2$ of eq.~(\ref{eq:Ct}), on the $32^3 \times 16$ lattice with lattice spacing $a$=0.064 fm.}
\label{tab:Ct_ud_b620}
\end{table*}

\begin{table*}[tbp]
\centering
\begin{adjustbox}{max width=\linewidth}
\begin{tabular}{|c|c|c|c|c|c|c|c|}
\hline
$a[\text{fm}]$ & $T_r'[\text{MeV}]$ & $\chi_P$ & $\chi_S$ & $\chi_V$ & $\chi_A$ & $\chi_T$ & $\chi_X$ \\
\hline
0.075 & 715 & 4.285565e-2 $\pm$ 9.8e-5& 4.28556e-2 $\pm$ 9.8e-5& 3.59551e-2 $\pm$ 8.55e-5& 3.595505e-2 $\pm$ 8.55e-5& 3.02276e-2 $\pm$ 8.05e-5& 3.022755e-2 $\pm$ 8.05e-5 \\
0.069 & 715 & 4.313945e-2 $\pm$ 7.8e-5& 4.313945e-2 $\pm$ 7.8e-5& 3.57104e-2 $\pm$ 7.5e-5& 3.57104e-2 $\pm$ 7.5e-5& 2.961455e-2 $\pm$ 7.3e-5& 2.961455e-2 $\pm$ 7.3e-5 \\
0.064 & 715 & 4.373065e-2 $\pm$ 5.55e-5& 4.373065e-2 $\pm$ 5.55e-5& 3.561965e-2 $\pm$ 4.82e-5& 3.561965e-2 $\pm$ 4.82e-5& 2.90647e-2 $\pm$ 4.505e-5& 2.90647e-2 $\pm$ 4.505e-5 \\
\hline
0.075 & 770 & 4.26956e-2 $\pm$ 8.95e-5& 4.26956e-2 $\pm$ 8.95e-5& 3.63125e-2 $\pm$ 7.95e-5& 3.63125e-2 $\pm$ 7.95e-5& 3.09565e-2 $\pm$ 7.55e-5& 3.095645e-2 $\pm$ 7.55e-5 \\
0.069 & 770 & 4.29982e-2 $\pm$ 7.2e-5& 4.299815e-2 $\pm$ 7.2e-5& 3.60414e-2 $\pm$ 7.05e-5& 3.604135e-2 $\pm$ 7.05e-5& 3.02733e-2 $\pm$ 6.9e-5& 3.02733e-2 $\pm$ 6.9e-5 \\
0.064 & 770 & 4.32572e-2 $\pm$ 6.35e-5& 4.32572e-2 $\pm$ 6.35e-5& 3.58288e-2 $\pm$ 5.8e-5& 3.58288e-2 $\pm$ 5.8e-5& 2.973125e-2 $\pm$ 5.65e-5& 2.973125e-2 $\pm$ 5.65e-5 \\
\hline
\end{tabular}
\end{adjustbox}
\caption{The bare susceptibilities (\ref{eq:chi_bare_lat}) of each channel at the two additional
reference temperatures $T_r'=715$ and $770$~MeV, used together with table~\ref{tab:chi_Tr} for the
reference-independence check of table~\ref{tab:Tr_compare}. Each of the three references coincides
with $aT_r=1/4$ ($N_t=4$) on one lattice spacing and is interpolated on the other two: $658$~MeV is
direct at $\beta=6.15$, $715$~MeV at $\beta=6.18$, and $770$~MeV at $\beta=6.20$. As at $T_r=658$~MeV, the chiral
partners agree within uncertainties at all three lattice spacings.}
\label{tab:chi_Tr_extra}
\end{table*}

\begin{table*}[tbp]
\centering
\begin{adjustbox}{max width=\linewidth}
\begin{tabular}{|c|c|c|c|c|c|c|c|}
\hline
$a[\text{fm}]$ & $T_r[\text{MeV}]$ & $\chi_S$ & $\chi_P$ & $\chi_V$ & $\chi_A$ & $\chi_T$ & $\chi_X$ \\
\hline
0.075 & 658 & 4.30237e-2 $\pm$ 1.075e-4& 4.302375e-2 $\pm$ 1.075e-4& 3.55848e-2 $\pm$ 9.25e-5& 3.558475e-2 $\pm$ 9.25e-5& 2.94832e-2 $\pm$ 8.55e-5& 2.948315e-2 $\pm$ 8.55e-5 \\
0.069 & 658 & 4.36669e-2 $\pm$ 6.35e-5& 4.366695e-2 $\pm$ 6.35e-5& 3.547735e-2 $\pm$ 5.9e-5& 3.547735e-2 $\pm$ 5.9e-5& 2.887425e-2 $\pm$ 5.6e-5& 2.887425e-2 $\pm$ 5.6e-5 \\
0.064 & 658 & 4.423165e-2 $\pm$ 6e-5& 4.42317e-2 $\pm$ 6e-5& 3.54021e-2 $\pm$ 4.7e-5& 3.540205e-2 $\pm$ 4.7e-5& 2.838315e-2 $\pm$ 3.92e-5& 2.838315e-2 $\pm$ 3.92e-5 \\
\hline
\end{tabular}
\end{adjustbox}
\caption{The bare susceptibilities (\ref{eq:chi_bare_lat}) of each channel at the common
physical reference temperature $T_r = 658$~MeV, listed separately for the two members of each
chiral pair so that the degeneracies $\chi_S \simeq \chi_P$, $\chi_V \simeq \chi_A$ and
$\chi_T \simeq \chi_X$ used in the temperature subtraction can be verified directly. At $\beta=6.15$
($a=0.075$~fm) this temperature coincides with $aT_r=1/4$ ($N_t=4$). At $\beta=6.18$ and $6.20$ it is
obtained by interpolation between the $N_t=6$ and $N_t=4$ ensembles. The partners agree within
uncertainties at all three lattice spacings.}
\label{tab:chi_Tr}
\end{table*}

\begin{table*}[tbp]
\setlength{\tabcolsep}{5pt}
\centering
\small
\begin{adjustbox}{max width=\linewidth}
\begin{tabular}{|c|c|c|c|c|}
\hline
$T[\text{MeV}]$ & $a[\text{fm}]$ & $\chi_S^\reg$ & $\chi_P^\reg$ & $\kappa_{PS}$ \\
\hline
164 & 0.075 & 9.913152e-03 $\pm$ 4.0575e-04& 1.0354935e-02 $\pm$ 3.994e-04& 2.17970e-02 $\pm$ 9.5175e-03 \\
179 & 0.069 & 8.3175035e-03 $\pm$ 3.4895e-04& 8.3808155e-03 $\pm$ 3.577e-04& 3.79150e-03 $\pm$ 1.7975e-03 \\
192 & 0.064 & 7.129055e-03 $\pm$ 1.6505e-04& 7.1627925e-03 $\pm$ 1.652e-04& 2.36061e-03 $\pm$ 7.1569e-04 \\
219 & 0.075 & 7.216618e-03 $\pm$ 2.5915e-04& 7.28522e-03 $\pm$ 2.666e-04& 4.73057e-03 $\pm$ 1.3948e-03 \\
238 & 0.069 & 6.234774e-03 $\pm$ 1.558e-04& 6.2445585e-03 $\pm$ 1.5585e-04& 7.84062e-04 $\pm$ 1.8777e-04 \\
257 & 0.064 & 5.4073395e-03 $\pm$ 1.6305e-04& 5.4099045e-03 $\pm$ 1.6305e-04& 2.37127e-04 $\pm$ 5.7594e-05 \\
263 & 0.075 & 5.0865675e-03 $\pm$ 1.6715e-04& 5.0927375e-03 $\pm$ 1.6705e-04& 6.06160e-04 $\pm$ 1.6347e-04 \\
286 & 0.069 & 4.59560975e-03 $\pm$ 1.303e-04& 4.59742855e-03 $\pm$ 1.303e-04& 1.97845e-04 $\pm$ 5.2926e-05 \\
308 & 0.064 & 4.1404347e-03 $\pm$ 1.467e-04& 4.14124015e-03 $\pm$ 1.466e-04& 9.72575e-05 $\pm$ 3.1563e-05 \\
329 & 0.075 & 4.08617125e-03 $\pm$ 1.735e-04& 4.0893186e-03 $\pm$ 1.7335e-04& 3.84974e-04 $\pm$ 7.9371e-05 \\
357 & 0.069 & 3.3400626e-03 $\pm$ 1.7455e-04& 3.34040475e-03 $\pm$ 1.7455e-04& 5.12218e-05 $\pm$ 1.5589e-05 \\
385 & 0.064 & 3.05594965e-03 $\pm$ 1.2855e-04& 3.05617035e-03 $\pm$ 1.286e-04& 3.61037e-05 $\pm$ 9.3457e-06 \\
\hline
\end{tabular}
\end{adjustbox}
\caption{Regularized scalar and pseudoscalar susceptibilities $\chi_S^\reg$, $\chi_P^\reg$ and the RGI symmetry ratio $\kappa_{PS}$ for three lattice spacings and twelve temperatures, obtained with the common physical reference temperature $T_r=658$~MeV subtracted at every lattice spacing (Sec.~\ref{subsec:results_kappa}). The susceptibilities are quoted to eight significant figures so that $\kappa_{PS}$ can be reconstructed from $\chi_S^\reg$ and $\chi_P^\reg$ to well within its jackknife uncertainty (see text).}
\label{tab:chi_PS_ud}
\end{table*}

\begin{table*}[tbp]
\setlength{\tabcolsep}{5pt}
\centering
\small
\begin{adjustbox}{max width=\linewidth}
\begin{tabular}{|c|c|c|c|c|}
\hline
$T[\text{MeV}]$ & $a[\text{fm}]$ & $\chi_V^\reg$ & $\chi_A^\reg$ & $\kappa_{VA}$ \\
\hline
164 & 0.075 & -1.3620404e-03 $\pm$ 1.3025e-04& -1.3635762e-03 $\pm$ 1.303e-04& 5.63466e-04 $\pm$ 6.5492e-05 \\
179 & 0.069 & -9.0980795e-04 $\pm$ 9.925e-05& -9.1069305e-04 $\pm$ 9.925e-05& 4.86194e-04 $\pm$ 6.7094e-05 \\
192 & 0.064 & -8.454074e-04 $\pm$ 8.745e-05& -8.458042e-04 $\pm$ 8.745e-05& 2.34621e-04 $\pm$ 2.9567e-05 \\
219 & 0.075 & -1.40228575e-03 $\pm$ 1.2245e-04& -1.4029904e-03 $\pm$ 1.2245e-04& 2.51197e-04 $\pm$ 4.3978e-05 \\
238 & 0.069 & -9.9148545e-04 $\pm$ 8.76e-05& -9.9179145e-04 $\pm$ 8.76e-05& 1.54307e-04 $\pm$ 1.2758e-05 \\
257 & 0.064 & -7.8802745e-04 $\pm$ 7.975e-05& -7.881305e-04 $\pm$ 7.975e-05& 6.53771e-05 $\pm$ 7.0477e-06 \\
263 & 0.075 & -1.36663905e-03 $\pm$ 1.2045e-04& -1.36681035e-03 $\pm$ 1.2045e-04& 6.26647e-05 $\pm$ 4.1765e-06 \\
286 & 0.069 & -1.00970255e-03 $\pm$ 8.775e-05& -1.0098165e-03 $\pm$ 8.77e-05& 5.64368e-05 $\pm$ 4.1238e-06 \\
308 & 0.064 & -6.8054525e-04 $\pm$ 1.0045e-04& -6.8059425e-04 $\pm$ 1.0045e-04& 3.59943e-05 $\pm$ 4.6614e-06 \\
329 & 0.075 & -1.3725132e-03 $\pm$ 1.163e-04& -1.3726225e-03 $\pm$ 1.163e-04& 3.98131e-05 $\pm$ 2.4942e-06 \\
357 & 0.069 & -9.8207135e-04 $\pm$ 1.0855e-04& -9.8213055e-04 $\pm$ 1.0855e-04& 3.01441e-05 $\pm$ 3.2522e-06 \\
385 & 0.064 & -6.1642165e-04 $\pm$ 8.625e-05& -6.1644765e-04 $\pm$ 8.625e-05& 2.10762e-05 $\pm$ 2.7774e-06 \\
\hline
\end{tabular}
\end{adjustbox}
\caption{Regularized vector and axial-vector susceptibilities $\chi_V^\reg$, $\chi_A^\reg$ and the RGI symmetry ratio $\kappa_{VA}$ for three lattice spacings and twelve temperatures, obtained with the common physical reference temperature $T_r=658$~MeV subtracted at every lattice spacing (Sec.~\ref{subsec:results_kappa}). The susceptibilities are quoted to eight significant figures so that $\kappa_{VA}$ can be reconstructed from $\chi_V^\reg$ and $\chi_A^\reg$ to well within its jackknife uncertainty (see text).}
\label{tab:chi_VA_ud}
\end{table*}

\begin{table*}[tbp]
\setlength{\tabcolsep}{5pt}
\centering
\small
\begin{adjustbox}{max width=\linewidth}
\begin{tabular}{|c|c|c|c|c|}
\hline
$T[\text{MeV}]$ & $a[\text{fm}]$ & $\chi_T^\reg$ & $\chi_X^\reg$ & $\kappa_{TX}$ \\
\hline
164 & 0.075 & -4.6866373e-03 $\pm$ 1.0115e-04& -4.68679585e-03 $\pm$ 1.011e-04& 1.69138e-05 $\pm$ 8.9847e-05 \\
179 & 0.069 & -3.96614145e-03 $\pm$ 7.085e-05& -3.9668441e-03 $\pm$ 7.085e-05& 8.85709e-05 $\pm$ 2.3704e-05 \\
192 & 0.064 & -3.52213185e-03 $\pm$ 5.59e-05& -3.52244115e-03 $\pm$ 5.59e-05& 4.39034e-05 $\pm$ 1.4993e-05 \\
219 & 0.075 & -4.55587565e-03 $\pm$ 9.88e-05& -4.5562359e-03 $\pm$ 9.88e-05& 3.95350e-05 $\pm$ 6.7434e-06 \\
238 & 0.069 & -3.89697985e-03 $\pm$ 6.915e-05& -3.89718885e-03 $\pm$ 6.915e-05& 2.68159e-05 $\pm$ 1.7399e-06 \\
257 & 0.064 & -3.3090318e-03 $\pm$ 5.335e-05& -3.30910345e-03 $\pm$ 5.335e-05& 1.08282e-05 $\pm$ 8.6098e-07 \\
263 & 0.075 & -4.328279e-03 $\pm$ 9.585e-05& -4.3284153e-03 $\pm$ 9.585e-05& 1.57407e-05 $\pm$ 9.9988e-07 \\
286 & 0.069 & -3.62313555e-03 $\pm$ 6.555e-05& -3.6232271e-03 $\pm$ 6.555e-05& 1.26331e-05 $\pm$ 5.6017e-07 \\
308 & 0.064 & -3.01015325e-03 $\pm$ 6.445e-05& -3.01019235e-03 $\pm$ 6.445e-05& 6.49667e-06 $\pm$ 2.3957e-07 \\
329 & 0.075 & -4.1055431e-03 $\pm$ 9.665e-05& -4.1056303e-03 $\pm$ 9.665e-05& 1.06173e-05 $\pm$ 6.8054e-07 \\
357 & 0.069 & -3.27507635e-03 $\pm$ 7.68e-05& -3.27512595e-03 $\pm$ 7.68e-05& 7.57216e-06 $\pm$ 2.7448e-07 \\
385 & 0.064 & -2.5918733e-03 $\pm$ 5.885e-05& -2.5918948e-03 $\pm$ 5.885e-05& 4.14285e-06 $\pm$ 1.5337e-07 \\
\hline
\end{tabular}
\end{adjustbox}
\caption{Regularized tensor vector and axial-tensor vector susceptibilities $\chi_T^\reg$, $\chi_X^\reg$ and the RGI symmetry ratio $\kappa_{TX}$ for three lattice spacings and twelve temperatures, obtained with the common physical reference temperature $T_r=658$~MeV subtracted at every lattice spacing (Sec.~\ref{subsec:results_kappa}). The susceptibilities are quoted to eight significant figures so that $\kappa_{TX}$ can be reconstructed from $\chi_T^\reg$ and $\chi_X^\reg$ to well within its jackknife uncertainty (see text).}
\label{tab:chi_TX_ud}
\end{table*}


\begin{thebibliography}{99}
\bibitem{Nambu:1961tp}
Y.~Nambu and G.~Jona-Lasinio,
``Dynamical model of elementary particles based on an analogy with Superconductivity. 1.'',
\href{https://doi.org/10.1103/PhysRev.122.345}{\blue Phys. Rev. \textbf{122}, 345-358 (1961)}

\bibitem{Nambu:1961fr}
Y.~Nambu and G.~Jona-Lasinio,
``Dynamical model of elementary particles based on an analogy with superconductivity. II.'',
\href{https://doi.org/10.1103/PhysRev.124.246}{\blue Phys. Rev. \textbf{124}, 246-254 (1961)}

\bibitem{Adler:1969gk}
S.~L.~Adler,
``Axial vector vertex in spinor electrodynamics'',
\href{https://doi.org/10.1103/PhysRev.177.2426}{\blue Phys. Rev. \textbf{177}, 2426-2438 (1969)}

\bibitem{Bell:1969ts}
J.~S.~Bell and R.~Jackiw,
``A PCAC puzzle: $\pi^0 \to \gamma \gamma$ in the $\sigma$ model'',
\href{https://doi.org/10.1007/BF02823296}{\blue Nuovo Cim. A \textbf{60}, 47-61 (1969)}

\bibitem{Fujikawa:1979ay}
K.~Fujikawa,
``Path Integral Measure for Gauge Invariant Fermion Theories'',
\href{https://doi.org/10.1103/PhysRevLett.42.1195}{\blue Phys. Rev. Lett. \textbf{42}, 1195-1198 (1979)}

\bibitem{tHooft:1976snw}
G.~'t Hooft,
``Computation of the Quantum Effects Due to a Four-Dimensional Pseudoparticle'',
Phys. Rev. D \textbf{14}, 3432-3450 (1976)
\href{https://doi.org/10.1103/PhysRevD.14.3432}{\blue [erratum: Phys. Rev. D \textbf{18}, 2199 (1978)]}

\bibitem{Witten:1979vv}
E.~Witten,
``Current Algebra Theorems for the U(1) Goldstone Boson'',
\href{https://doi.org/10.1016/0550-3213(79)90031-2}{\blue Nucl. Phys. B \textbf{156}, 269-283 (1979)}

\bibitem{Veneziano:1979ec}
G.~Veneziano,
``U(1) Without Instantons'',
\href{https://doi.org/10.1016/0550-3213(79)90332-8}{\blue Nucl. Phys. B \textbf{159}, 213-224 (1979)}

\bibitem{Aoki:2006we}
Y.~Aoki, G.~Endrodi, Z.~Fodor, S.~D.~Katz and K.~K.~Szabo,
``The Order of the quantum chromodynamics transition predicted by the standard model of particle physics'', 
\href{https://doi.org/10.1038/nature05120}{\blue Nature \textbf{443}, 675-678 (2006)}
[arXiv:hep-lat/0611014 [hep-lat]].

\bibitem{Borsanyi:2013bia}
S.~Borsanyi, Z.~Fodor, C.~Hoelbling, S.~D.~Katz, S.~Krieg and K.~K.~Szabo,
``Full result for the QCD equation of state with 2+1 flavors'', 
\href{https://doi.org/10.1016/j.physletb.2014.01.007}{\blue Phys. Lett. B \textbf{730}, 99-104 (2014)}
[arXiv:1309.5258 [hep-lat]].

\bibitem{HotQCD:2014kol}
A.~Bazavov \textit{et al.} [HotQCD],
``Equation of state in ( 2+1 )-flavor QCD'', 
\href{https://doi.org/10.1103/PhysRevD.90.094503}{\blue Phys. Rev. D \textbf{90}, 094503 (2014)}
[arXiv:1407.6387 [hep-lat]].

\bibitem{HotQCD:2018pds}
A.~Bazavov \textit{et al.} [HotQCD],
``Chiral crossover in QCD at zero and non-zero chemical potentials'',
\href{https://doi.org/10.1016/j.physletb.2019.05.013}{\blue Phys. Lett. B \textbf{795}, 15-21 (2019)}
[arXiv:1812.08235 [hep-lat]].

\bibitem{Borsanyi:2020fev}
S.~Borsanyi, Z.~Fodor, J.~N.~Guenther, R.~Kara, S.~D.~Katz, P.~Parotto, A.~Pasztor, C.~Ratti and K.~K.~Szabo,
``QCD Crossover at Finite Chemical Potential from Lattice Simulations'',
\href{https://doi.org/10.1103/PhysRevLett.125.052001}{\blue Phys. Rev. Lett. \textbf{125}, no.5, 052001 (2020)}
[arXiv:2002.02821 [hep-lat]].

\bibitem{Pisarski:1983ms}
R.~D.~Pisarski and F.~Wilczek,
``Remarks on the Chiral Phase Transition in Chromodynamics'',
\href{https://doi.org/10.1103/PhysRevD.29.338}{\blue Phys. Rev. D \textbf{29}, 338-341 (1984)}

\bibitem{Leutwyler:1992yt}
H.~Leutwyler and A.~V.~Smilga,
``Spectrum of Dirac operator and role of winding number in QCD'',
\href{https://doi.org/10.1103/PhysRevD.46.5607}{\blue Phys. Rev. D \textbf{46}, 5607-5632 (1992)}

\bibitem{Cohen:1996ng}
T.~D.~Cohen,
``QCD inequalities, the high temperature phase of QCD, and $U(1)_A$ symmetry",
\href{https://doi.org/10.1103/PhysRevD.54.R1867}{\blue Phys. Rev. D \textbf{54}, R1867-R1870 (1996)}
[arXiv:hep-ph/9601216 [hep-ph]].

\bibitem{Aoki:2012yj}
S.~Aoki, H.~Fukaya and Y.~Taniguchi,
``Chiral symmetry restoration, eigenvalue density of Dirac operator and axial U(1) anomaly 
  at finite temperature'', 
  \href{https://doi.org/10.1103/PhysRevD.86.114512}{\blue Phys. Rev. D \textbf{86}, 114512 (2012)}
[arXiv:1209.2061 [hep-lat]].

\bibitem{Cossu:2013uua}
G.~Cossu, S.~Aoki, H.~Fukaya, S.~Hashimoto, T.~Kaneko, H.~Matsufuru and J.~I.~Noaki,
``Finite temperature study of the axial U(1) symmetry on the lattice with overlap fermion formulation,''
\href{https://doi.org/10.1103/PhysRevD.87.114514}{\blue Phys. Rev. D \textbf{87}, no.11, 114514 (2013)}
\href{https://doi.org/10.1103/PhysRevD.88.019901}{\blue [erratum: Phys. Rev. D \textbf{88}, no.1, 019901 (2013)]}
[arXiv:1304.6145 [hep-lat]].

\bibitem{Buchoff:2013nra}
M.~I.~Buchoff, M.~Cheng, N.~H.~Christ, H.~T.~Ding, C.~Jung, F.~Karsch, Z.~Lin, R.~D.~Mawhinney, 
S.~Mukherjee and P.~Petreczky, \textit{et al.}
``QCD chiral transition, U(1)A symmetry and the dirac spectrum using domain wall fermions'', 
\href{https://doi.org/10.1103/PhysRevD.89.054514}{\blue Phys. Rev. D \textbf{89}, no.5, 054514 (2014)}
[arXiv:1309.4149 [hep-lat]].

\bibitem{Brandt:2016daq}
B.~B.~Brandt, A.~Francis, H.~B.~Meyer, O.~Philipsen, D.~Robaina and H.~Wittig,
``On the strength of the $U_A(1)$ anomaly at the chiral phase transition in $N_f=2$ QCD'',
\href{https://doi.org/10.1007/JHEP12(2016)158}{\blue 
JHEP \textbf{12}, 158 (2016)}
[arXiv:1608.06882 [hep-lat]].

\bibitem{Tomiya:2016jwr}
A.~Tomiya, G.~Cossu, S.~Aoki, H.~Fukaya, S.~Hashimoto, T.~Kaneko and J.~Noaki,
``Evidence of effective axial U(1) symmetry restoration at high temperature QCD'',
\href{https://doi.org/10.1103/PhysRevD.96.034509}{\blue Phys. Rev. D \textbf{96}, no.3, 034509 (2017)}
\href{https://doi.org/10.1103/PhysRevD.96.079902}{\blue [addendum: Phys. Rev. D \textbf{96}, 079902 (2017)]}
[arXiv:1612.01908 [hep-lat]].

\bibitem{Ding:2020xlj}
H.~T.~Ding, S.~T.~Li, S.~Mukherjee, A.~Tomiya, X.~D.~Wang and Y.~Zhang,
``Correlated Dirac Eigenvalues and Axial Anomaly in Chiral Symmetric QCD'',
\href{https://doi.org/10.1103/PhysRevLett.126.082001}{\blue Phys. Rev. Lett. \textbf{126}, no.8, 082001 (2021)}
[arXiv:2010.14836 [hep-lat]].

\bibitem{Kaczmarek:2023bxb}
O.~Kaczmarek, R.~Shanker and S.~Sharma,
``Eigenvalues of the QCD Dirac matrix with improved staggered quarks in the continuum limit'',
\href{https://doi.org/10.1103/PhysRevD.108.094501}{\blue Phys. Rev. D \textbf{108}, no.9, 094501 (2023)}
[arXiv:2301.11610 [hep-lat]].

\bibitem{Aoki:2020noz}
S.~Aoki \textit{et al.} [JLQCD],
``Study of the axial $U(1)$ anomaly at high temperature with lattice chiral fermions'',
\href{https://doi.org/10.1103/PhysRevD.103.074506}{\blue Phys. Rev. D \textbf{103}, no.7, 074506 (2021)}
[arXiv:2011.01499 [hep-lat]].

\bibitem{Chiu:2023hnm}
T.~W.~Chiu,
``Symmetries of meson correlators in high-temperature QCD with physical (u/d,s,c) domain-wall quarks'',
\href{https://doi.org/10.1103/PhysRevD.107.114501}{\blue Phys. Rev. D \textbf{107}, no.11, 114501 (2023)}
[arXiv:2302.06073 [hep-lat]].

\bibitem{Gavai:2024mcj}
R.~V.~Gavai, M.~E.~Jaensch, O.~Kaczmarek, F.~Karsch, M.~Sarkar, R.~Shanker, S.~Sharma, S.~Sharma 
and T.~Ueding,
``Aspects of the chiral crossover transition in (2+1)-flavor QCD with M{\"o}bius domain-wall fermions'',
\href{https://doi.org/10.1103/PhysRevD.111.034507}{\blue Phys. Rev. D \textbf{111}, no.3, 034507 (2025)}
[arXiv:2411.10217 [hep-lat]].

\bibitem{Ding:2026gao}
H.~T.~Ding,
``Lattice QCD at finite temperature and density,''
\href{https://doi.org/10.22323/1.518.0016}{\blue PoS \textbf{LATTICE2025}, 016 (2026)}
[arXiv:2603.16230 [hep-lat]].

\bibitem{Kaplan:1992bt}
D.~B.~Kaplan,
``A Method for simulating chiral fermions on the lattice'',
\href{https://doi.org/10.1016/0370-2693(92)91112-M}{\blue Phys. Lett. B \textbf{288}, 342-347 (1992)}
[arXiv:hep-lat/9206013 [hep-lat]].

\bibitem{Kaplan:1992sg}
D.~B.~Kaplan,
``Chiral fermions on the lattice'',
\href{https://doi.org/10.1016/0920-5632(93)90282-B}
     {\blue Nucl. Phys. B Proc. Suppl. \textbf{30}, 597-600 (1993)}

\bibitem{Neuberger:1997fp}
H.~Neuberger,
``Exactly massless quarks on the lattice'',
\href{https://doi.org/10.1016/S0370-2693(97)01368-3}{\blue Phys. Lett. B \textbf{417}, 141-144 (1998)}
[arXiv:hep-lat/9707022 [hep-lat]].

\bibitem{Narayanan:1994gw}
R.~Narayanan and H.~Neuberger,
``A Construction of lattice chiral gauge theories'',
\href{https://doi.org/10.1016/0550-3213(95)00111-5}{\blue Nucl. Phys. B \textbf{443}, 305-385 (1995)}
[arXiv:hep-th/9411108 [hep-th]].

\bibitem{Chiu:2026upk}
T.~W.~Chiu,
``Renormalization of meson susceptibilities and RG-invariant symmetry ratios in QCD,''
[arXiv:2607.17816 [hep-lat]]; some of the results cited here appear in the revised version,
which is under review and will be posted on arXiv.

\bibitem{Chiu:2024jyz}
T.~W.~Chiu,
``Symmetries of spatial correlators of light and heavy mesons in high temperature lattice QCD'',
\href{https://doi.org/10.1103/PhysRevD.110.014502}{\blue Phys. Rev. D \textbf{110}, no.1, 014502 (2024)}
[arXiv:2404.15932 [hep-lat]].

\bibitem{Chiu:2024bqx}
T.~W.~Chiu,
``Symmetries in High-Temperature Lattice QCD with (u, d, s, c, b) Optimal Domain-Wall Quarks'',
\href{https://doi.org/10.3390/sym17050700}{\blue Symmetry \textbf{17}, no.5, 700 (2025)}
[arXiv:2411.16705 [hep-lat]].

\bibitem{CohenJi:1997}
T.~D.~Cohen and X.~D.~Ji,
``Chiral multiplets of hadron currents'',
\href{https://doi.org/10.1103/PhysRevD.55.6870}{\blue Phys. Rev. D \textbf{55}, 6870-6876 (1997)}
[arXiv:hep-ph/9612302 [hep-ph]].

\bibitem{Dolan:1973qd}
L.~Dolan and R.~Jackiw,
``Symmetry Behavior at Finite Temperature,''
\href{https://doi.org/10.1103/PhysRevD.9.3320}{\blue Phys. Rev. D \textbf{9}, 3320-3341 (1974)}

\bibitem{Weinberg:1974hy}
S.~Weinberg,
``Gauge and Global Symmetries at High Temperature,''
\href{https://doi.org/10.1103/PhysRevD.9.3357}{\blue Phys. Rev. D \textbf{9}, 3357-3378 (1974)}

\bibitem{Bernard:1974bq}
C.~W.~Bernard,
``Feynman Rules for Gauge Theories at Finite Temperature,''
\href{https://doi.org/10.1103/PhysRevD.9.3312}{\blue Phys. Rev. D \textbf{9}, 3312-3319 (1974)}

\bibitem{Kislinger:1975ab}
M.~B.~Kislinger and P.~D.~Morley,
``Collective Phenomena in Gauge Theories. 2. Renormalization in Finite Temperature Field Theory,''
\href{https://10.1103/PhysRevD.13.2771}{\blue Phys. Rev. D \textbf{13}, 2771 (1976)}

\bibitem{Landsman:1986uw}
N.~P.~Landsman and C.~G.~van Weert,
``Real and Imaginary Time Field Theory at Finite Temperature and Density,''
\href{https://doi.org/10.1016/0370-1573(87)90121-9}{\blue Phys. Rept. \textbf{145}, 141 (1987)}

\bibitem{Chiu:2002ir}
  T.~W.~Chiu,
``Optimal domain wall fermions'', 
\href{http://doi.org/10.1103/PhysRevLett.90.071601}{\blue Phys.\ Rev.\ Lett.\  {\bf 90}, 071601 (2003)}
  [hep-lat/0209153];
  

\bibitem{Chiu:2015sea}
  T.~W.~Chiu,
``Domain-Wall Fermion with $ R_5 $ Symmetry'',
\href{https://doi.org/10.1016/j.physletb.2015.03.036}{\blue Phys.\ Lett.\ B {\bf 744}, 95 (2015)}
  [arXiv:1503.01750 [hep-lat]].

\bibitem{Wilson:1974sk}
K.~G.~Wilson,
``Confinement of Quarks'',
\href{https://doi.org/10.1103/PhysRevD.10.2445}{\blue Phys. Rev. D \textbf{10}, 2445-2459 (1974)}

\bibitem{Chiu:2011bm}
  T.~W.~Chiu, T.~H.~Hsieh, Y.~Y.~Mao [TWQCD Collaboration],
``Pseudoscalar Meson in Two Flavors QCD with the Optimal Domain-Wall Fermion'',
\href{https://doi.org/10.1016/j.physletb.2012.09.067}{\blue Phys.\ Lett.\ B {\bf 717}, 420 (2012)}
  [arXiv:1109.3675 [hep-lat]].

\bibitem{Chen:2014hyy}
  Y.~C.~Chen, T.~W.~Chiu [TWQCD Collaboration],
``Exact Pseudofermion Action for Monte Carlo Simulation of Domain-Wall Fermion'',
\href{https://doi.org/10.1016/j.physletb.2014.09.016}{\blue Phys.\ Lett.\ B {\bf 738}, 55 (2014)}
  [arXiv:1403.1683 [hep-lat]].

\bibitem{Chen:2022fid}
Y.~C.~Chen, T.~W.~Chiu and T.~H.~Hsieh [TWQCD Collaboration],
``Topological susceptibility in finite temperature QCD with physical (u/d,s,c) domain-wall quarks'', 
\href{https://doi.org/10.1103/PhysRevD.106.074501}{\blue Phys. Rev. D \textbf{106}, no.7, 074501 (2022)}
[arXiv:2204.01556 [hep-lat]].

\bibitem{Chen:2012jya}
  Y.~C.~Chen, T.~W.~Chiu [TWQCD Collaboration],
``Chiral Symmetry and the Residual Mass in Lattice QCD with the Optimal Domain-Wall Fermion'',
\href{https://doi.org/10.1103/PhysRevD.86.094508}{\blue Phys.\ Rev.\ D {\bf 86}, 094508 (2012)}
  [arXiv:1205.6151 [hep-lat]].

\bibitem{Narayanan:2006rf}
  R.~Narayanan and H.~Neuberger,
``Infinite N phase transitions in continuum Wilson loop operators'',
\href{https://doi.org/10.1088/1126-6708/2006/03/064}{\blue JHEP {\bf 0603}, 064 (2006)}
  [hep-th/0601210].

\bibitem{Luscher:2010iy}
  M.~Luscher,
``Properties and uses of the Wilson flow in lattice QCD'',
\href{https://doi.org/10.1007/JHEP08(2010)071}{\blue JHEP {\bf 1008}, 071 (2010)}; 
 Erratum: [\href{https://doi.org/10.1007/JHEP03(2014)092}{\blue JHEP {\bf 1403}, 092 (2014)}]
  [arXiv:1006.4518 [hep-lat]].

\bibitem{Bazavov:2015yea}
  A.~Bazavov {\it et al.} [MILC Collaboration],
``Gradient flow and scale setting on MILC HISQ ensembles'',
\href{https://doi.org/10.1103/PhysRevD.93.094510}{\blue Phys.\ Rev.\ D {\bf 93}, no. 9, 094510 (2016)}
  [arXiv:1503.02769 [hep-lat]].

\bibitem{Chiu:1998eu}
T.~W.~Chiu,
``Ginsparg-Wilson fermion propagators and chiral condensate'',
\href{https://doi.org/10.1103/PhysRevD.60.034503}{\blue Phys. Rev. D \textbf{60}, 034503 (1999)} 
[arXiv:hep-lat/9810052 [hep-lat]].

\bibitem{Chiu:inprep}
T.~W.~Chiu, \emph{et al.}, papers in preparation.

\bibitem{Bazavov:2012qja}
A.~Bazavov \textit{et al.} [HotQCD],
``The chiral transition and $U(1)_A$ symmetry restoration from lattice QCD using Domain Wall Fermions'',
\href{https://doi.org/10.1103/PhysRevD.86.094503}{\blue Phys. Rev. D \textbf{86}, 094503 (2012)}
[arXiv:1205.3535 [hep-lat]].

\bibitem{Petreczky:2016vrs}
P.~Petreczky, H.~P.~Schadler and S.~Sharma,
``The topological susceptibility in finite temperature QCD and axion cosmology'',
\href{https://doi.org/10.1016/j.physletb.2016.09.063}{\blue Phys. Lett. B \textbf{762}, 498-505 (2016)}
[arXiv:1606.03145 [hep-lat]].

\bibitem{Borsanyi:2016ksw}
S.~Borsanyi, Z.~Fodor, J.~Guenther, K.~H.~Kampert, S.~D.~Katz, T.~Kawanai, T.~G.~Kovacs, S.~W.~Mages, A.~Pasztor and F.~Pittler, \textit{et al.}
``Calculation of the axion mass based on high-temperature lattice quantum chromodynamics'', 
\href{https://doi.org/10.1038/nature20115}{\blue Nature \textbf{539}, no.7627, 69-71 (2016)}
[arXiv:1606.07494 [hep-lat]].

\bibitem{Athenodorou:2022aay}
A.~Athenodorou, C.~Bonanno, C.~Bonati, G.~Clemente, F.~D'Angelo, M.~D'Elia, L.~Maio, G.~Martinelli, F.~Sanfilippo and A.~Todaro,
``Topological susceptibility of N$_{f}$ = 2 + 1 QCD from staggered fermions spectral projectors at high temperatures'', \href{https://doi.org/10.1007/JHEP10(2022)197}{\blue JHEP \textbf{10}, 197 (2022)}
[arXiv:2208.08921 [hep-lat]].

\bibitem{Kotov:2025ilm}
A.~Y.~Kotov, M.~P.~Lombardo and A.~Trunin,
``Topological observables and {\ensuremath{\theta}} dependence in high temperature QCD 
from lattice simulations'', \href{https://doi.org/10.1007/JHEP09(2025)045}{\blue JHEP \textbf{09}, 045 (2025)}
[arXiv:2502.15407 [hep-lat]].

\bibitem{Gross:1980br}
D.~J.~Gross, R.~D.~Pisarski and L.~G.~Yaffe,
``QCD and Instantons at Finite Temperature'',
\href{https://doi.org/10.1103/RevModPhys.53.43}{\blue Rev. Mod. Phys. \textbf{53}, 43 (1981)}

\bibitem{Mao:2009sy}
Y.~Y.~Mao and T.~W.~Chiu [TWQCD Collaboration],
``Topological Susceptibility to the One-Loop Order in Chiral Perturbation Theory,''
\href{https://doi.org/10.1103/PhysRevD.80.034502}{\blue Phys. Rev. D \textbf{80}, 034502 (2009)}
[arXiv:0903.2146 [hep-lat]].

\bibitem{Glozman:2014mka}
L.~Y.~Glozman,
``SU(4) symmetry of the dynamical QCD string and genesis of hadron spectra'',
\href{https://doi.org/10.1140/epja/i2015-15027-x}{\blue Eur. Phys. J. A \textbf{51}, no.3, 27 (2015)}
[arXiv:1407.2798 [hep-ph]].

\bibitem{Rohrhofer:2019qwq}
C.~Rohrhofer, Y.~Aoki, G.~Cossu, H.~Fukaya, C.~Gattringer, L.~Y.~Glozman, S.~Hashimoto, 
C.~B.~Lang and S.~Prelovsek,
``Symmetries of spatial meson correlators in high temperature QCD'',
\href{https://doi.org/10.1103/PhysRevD.100.014502}{\blue Phys. Rev. D \textbf{100}, no.1, 014502 (2019)}
[arXiv:1902.03191 [hep-lat]].

\bibitem{Rohrhofer:2019qal}
C.~Rohrhofer, Y.~Aoki, L.~Y.~Glozman and S.~Hashimoto,
``Chiral-spin symmetry of the meson spectral function above $T_c$'',
\href{https://doi.org/10.1016/j.physletb.2020.135245}{\blue Phys. Lett. B \textbf{802}, 135245 (2020)}
[arXiv:1909.00927 [hep-lat]].

\bibitem{Alexandru:2019gdm}
A.~Alexandru and I.~Horv{\'a}th,
``Possible New Phase of Thermal QCD,''
\href{https://doi.org/10.1103/PhysRevD.100.094507}{\blue Phys. Rev. D \textbf{100}, no.9, 094507 (2019)}
[arXiv:1906.08047 [hep-lat]].

\bibitem{Meng:2023nxf}
X.~L.~Meng \textit{et al.} [{\ensuremath{\chi}}QCD and CLQCD],
``Separation of infrared and bulk in thermal QCD'', 
\href{https://doi.org/10.1007/JHEP12(2024)101}{\blue JHEP \textbf{12}, 101 (2024)}
[arXiv:2305.09459 [hep-lat]].

\bibitem{tHooft:1977nqb}
G.~'t Hooft,
``On the Phase Transition Towards Permanent Quark Confinement,''
\href{https://doi.org/10.1016/0550-3213(78)90153-0}{\blue Nucl. Phys. B \textbf{138}, 1-25 (1978)}

\bibitem{Mickley:2024vkm}
J.~A.~Mickley, C.~Allton, R.~Bignell and D.~B.~Leinweber,
``Center vortex evidence for a second finite-temperature QCD transition'', 
\href{https://doi.org/10.1103/PhysRevD.111.034508}{\blue Phys. Rev. D \textbf{111}, no.3, 034508 (2025)}
[arXiv:2411.19446 [hep-lat]].

\bibitem{Pelissetto:2013hqa}
A.~Pelissetto and E.~Vicari,
``Relevance of the axial anomaly at the finite-temperature chiral transition in QCD'',
\href{https://doi.org/10.1103/PhysRevD.88.105018}{\blue Phys. Rev. D \textbf{88}, no.10, 105018 (2013)}
[arXiv:1309.5446 [hep-lat]].

\bibitem{Kotov:2021rah}
A.~Y.~Kotov, M.~P.~Lombardo and A.~Trunin,
``QCD transition at the physical point, and its scaling window from twisted mass Wilson fermions,''
\href{http:/doi.org/10.1016/j.physletb.2021.136749}{\blue Phys. Lett. B \textbf{823}, 136749 (2021)}
[arXiv:2105.09842 [hep-lat]].

\bibitem{Klinger:2025xxb}
J.~P.~Klinger, R.~Kaiser and O.~Philipsen,
``The order of the chiral phase transition in massless many-flavour lattice QCD,''
\href{http://doi.org/10.22323/1.466.0172}{\blue PoS \textbf{LATTICE2024}, 172 (2025)}
[arXiv:2501.19251 [hep-lat]].

\bibitem{Braun:2020ada}
J.~Braun, W.~j.~Fu, J.~M.~Pawlowski, F.~Rennecke, D.~Rosenbl{\"u}h and S.~Yin,
``Chiral susceptibility in ( 2+1 )-flavor QCD'',
\href{https://doi.org/10.1103/PhysRevD.102.056010}{\blue Phys. Rev. D \textbf{102}, no.5, 056010 (2020)}
[arXiv:2003.13112 [hep-ph]].

\end{thebibliography}
\end{document}